\documentclass[aps,prb,floatfix,twocolumn,showpacs,superscriptaddress]{revtex4-1}  
\usepackage{graphicx}  
\usepackage{amsmath,amssymb}

\usepackage{physics}

\usepackage{flushend}
\usepackage[colorlinks = true,
            linkcolor = blue,
            urlcolor  = blue,
            citecolor = blue,
            anchorcolor = blue]{hyperref}
\usepackage{array}
\usepackage{float}
\usepackage{natbib}
\usepackage{booktabs}
\mathchardef\Re="023C
\mathchardef\Im="023D

\usepackage{color}
\usepackage{times}

\begin{document}

\title{Non-Hermiticity induced Exceptional Points and Skin Effect in the Dice-Haldane Model}

\author{Ronika Sarkar} \affiliation{Department of Physics,
Indian Institute of Science, Bangalore 560012, India.}
\author{Arka Bandyopadhyay}\affiliation{Solid State and Structural Chemistry Unit,
Indian Institute of Science, Bangalore 560012, India.}
\author{Awadhesh Narayan} \email[]{awadhesh@iisc.ac.in} \affiliation{Solid State and Structural Chemistry Unit,
Indian Institute of Science, Bangalore 560012, India.}

\vskip 0.25cm
\begin{abstract}

The interplay of topology and non-Hermiticity has led to diverse, exciting manifestations in a plethora of systems. In this work, we systematically investigate the role of non-Hermiticity in the Chern insulating Haldane model on a dice lattice. Due to the presence of a non-dispersive flat band, the dice-Haldane model hosts a topologically rich phase diagram with the non-trivial phases accommodating Chern numbers $\pm 2$. We introduce non-Hermiticity into this model in two ways -- through balanced non-Hermitian gain and loss, and by non-reciprocal hopping in one direction. Both these types of non-Hermiticity induce higher-order exceptional points of order three. Remarkably, the exceptional points at high symmetry points occur at odd integer values of the non-Hermiticity strength in the case of balanced gain and loss, and at odd integer multiples of $1/\sqrt{2}$ for non-reciprocal hopping. We substantiate the presence and the order of these higher-order exceptional points using the phase rigidity and its scaling. Further, we construct a phase diagram to identify and locate the occurrence of these exceptional points in the parameter space. Non-Hermiticity has yet more interesting consequences on a finite-sized lattice. Unlike for balanced gain and loss, in the case of non-reciprocal hopping, the nearest-neighbour dice lattice system under periodic boundary conditions accommodates a finite, non-zero spectral area in the complex plane. This manifests as the non-Hermitian skin effect when open boundary conditions are invoked. In the more general case of the dice-Haldane lattice model, the non-Hermitian skin effect can be caused by both gain and loss or non-reciprocity. Fascinatingly, the direction of localization of the eigenstates depends on the nature and strength of the non-Hermiticity. We establish the occurrence of the skin effect using the local density of states, inverse participation ratio and the edge probability, and demonstrate its robustness to disorder. Our results place the dice-Haldane model as an exciting platform to explore non-Hermitian physics.

\end{abstract}

\maketitle

\date{\today}

\section{Introduction}

In condensed matter physics, most of the intricate phases of matter, including magnetic and superconducting states, can be understood in the framework of the celebrated Landau theory \cite{sachdev_2011}. However, the two-dimensional electron gas at very low temperatures and under a strong transverse magnetic field exhibits a quantized Hall conductance \cite{PhysRevLett.45.494} -- such a quantization is not subject to any spontaneous symmetry breaking. Consequently, new concepts have been developed based on single-particle dynamics in topological band theory to unravel the advent of the integer quantum Hall effect. Haldane, in his seminal work, demonstrated that Dirac points in honeycomb lattices, such as graphene are protected by both inversion and time-reversal symmetry \cite{1,2,3,PhysRevB.78.075438,5,7}. Absence of any of these symmetries essentially leads to gapped spectra with distinct topological nature. In particular, Semenoff mass assigns an energy offset between the two sublattices of graphene and breaks the inversion symmetry \cite{semenoff1984condensed}. These inversion symmetry broken systems give rise to the normal or trivial insulators at half-filling. In contrast, a staggered magnetic field that turns the second nearest neighbor hoppings complex also breaks the time-reversal symmetry of the system without violating its translational symmetry. These time-reversal symmetry broken Chern insulators are at the heart of realizing quantized transverse Hall conductance in zero external magnetic field condition, namely the quantum anomalous Hall effect. In other words, the Haldane model is an elegant Chern insulator model on a honeycomb lattice, that allows tunability between topologically trivial and non-trivial phases by tuning the model parameters. The topological phase diagram of the Haldane model has successfully been realized in experiments using ultra cold fermionic atoms in optical lattices \cite{jotzu2014experimental}.

Moreover, unlike graphene, some of the bipartite lattices possess an unequal number of sublattices that offer an intriguing platform to realize perfectly or compact localized states. These compact localized states exhibit non-dispersive flat bands, i.e., the energy is independent of momentum in the electronic band structure. The underlying mechanism behind such flat bands can be well-explained in terms of destructive interference through various network paths. For example, the bipartite dice lattice \cite{sutherland1986localization,vidal1998aharonov,vidal2001disorder,bercioux2009massless,wang2011nearly,moller2012correlated,zhang2020compact,cheng2020predicting,wang2021flat,iurov2022optically} is one of the first and most prominent examples where such flat band physics was introduced. In a dice lattice, atoms are not only placed at the vertices of hexagons but also at the centers. Therefore, the number of sites with coordination number three is twice of those with coordination number six. In contrast to usual honeycomb lattices, three-component fermions invariably govern the low energy spectrum of the dice lattice. The two dispersive bands form Dirac cones and touch each other at symmetry points $K$ and $K^\prime$ of the Brillouin zone (BZ), while the remaining one is flat and lies at the Fermi level. The flat bands in the bipartite lattices occur because of the chiral symmetry. In other words, the bipartite model systems (such as dice) with a majority of one kind of sublattice invariably exhibit chiral flat bands \cite{sutherland1986localization}. The flat bands have been recently experimentally realized in photonic crystals employing ultrafast laser technology \cite{PhysRevLett.114.245504,PhysRevLett.114.245503,zong2016observation,PhysRevLett.121.075502}. Motivated by these interesting properties of the dice lattice, the Haldane model has been extended in the form of a three-band model with broken inversion and time-reversal symmetry \cite{21}. As expected, the topological phase diagram of the Haldane dice lattice is richer compared to that of the graphene with more interesting phases both within and outside the topologically non-trivial region \cite{23}.

Non-Hermitian physics \cite{25,26,27,28,30,31,32}, on the other hand, is a topic of growing widespread interest. Non-Hermiticity finds applicability in various fields of photonics, optics, and electronics, among others \cite{33,34,35,36,37,38,39}. Since these are open systems, the corresponding non-Hermitian Hamiltonian can accommodate for the gain and loss of particles or energy. Unusual properties, such as complex band spectra and non-orthogonal eigenstates, are the outcomes of such non-Hermitian Hamiltonians \cite{PhysRevLett.80.5243}. 
In particular, non-Hermitian systems can show a distinct class of spectral degeneracies known as exceptional points (EPs) \cite{40,41,42,43,44,banerjee2020controlling,banerjee2021non,45,46,47,48,49}, as well as exceptional contours \cite{chowdhury2021light,chowdhury2022exceptional}. At an EP the eigenvalues and the eigenvectors simultaneously coalesce making the Hamiltonian defective, i.e., non-diagonalizable. The number of eigenstates undergoing coalescence determines the order of the EP. The study of EPs has gained immense interest in the field of photonic systems \cite{parto2021non,midya2018non} and microwave cavities \cite{PhysRevLett.106.150403,dembowski2001experimental} among others, with interesting applications such as uni-directional sensitivity \cite{PhysRevLett.106.213901,hodaei2017enhanced}, laser mode selectivity \cite{feng2014single,hodaei2014parity} and opto-mechanical energy transfer \cite{xu2016topological}.
Non-Hermitian skin effect (NHSE), on the other hand, is a feature unique to non-Hermitian systems where a macroscopic fraction of eigenstates migrate to a boundary of the system as soon as open boundary conditions (OBC) are imposed \cite{yao2018edge, 51,52,53,54,55,56,58,60,borgnia2020non}. This extreme sensitivity of non-Hermitian systems to the boundary conditions leads to an anomalous bulk-boundary correspondence \cite{bulk-bound1,bulk-bound2,bulk-bound3,bulk-bound4}. The NHSE has been experimentally observed recently in photonic systems \cite{song2020two}, electrical circuits \cite{liu2021non,zou2021observation} and acoustic topological insulators \cite{zhang2021observation, zhang2021acoustic} among others.

In this paper, we systematically study the interplay between the effect of non-Hermiticity and different kinds of hopping terms in the dice lattice model. Later, we also study the role of disorder \cite{61,62,kramer1993localization,64,65,66,67,68,69,sarkar2022interplay} in the context of the non-trivial effects brought about by non-Hermiticity. We start with only the nearest neighbour hopping, then subsequently allow complex next-nearest neighbour hopping terms similar to the Haldane model, and finally, introduce the inversion breaking mass terms. For each case, we tune-up a non-Hermitian balanced gain and loss and investigate the changes in the eigenspectra and characterize the EPs which arise. We discover that third order EPs arise at odd integer values of the non-Hermiticity strength in each case. Their occurrence can be characterized using the phase rigidity which vanishes at the EP. Further, the scaling of phase rigidity with respect to the non-Hermiticity strength helps determine the order of the EP. When a non-reciprocal hopping is introduced instead of gain and loss, we find that third order EPs occur at odd integer multiples of $1/\sqrt{2}$. We also elucidate the complete phase diagram to determine the regions where such higher-order EPs can be found in the parameter space.
Non-reciprocal hopping has interesting consequences when we consider a finite sized dice-Haldane nanoribbon. For the dice lattice with only nearest neighbour coupling, under periodic boundary conditions (PBC), the spectrum under non-reciprocal hopping accommodates a finite, non-zero spectral area in the complex plane unlike the gain and loss case where the complex spectrum has an arc-like structure. This finite spectral area results in the occurrence of the NHSE when OBC are imposed on the lattice with non-reciprocal hopping. However, for the dice-Haldane model both gain and loss and nonreciprocal hopping exhibits finite spectral area under PBC and hence displays NHSE under OBC. The direction of the localization can be controlled by the choice of the non-Hermiticity and its strength.  We characterize the NHSE using the local density of states, inverse participation ratio (IPR) and the edge probability. This NHSE turns out to be fairly robust to disorder owing to its topological protection. However, at sufficiently large disorder strengths there is a complete destruction of NHSE accompanied by the bulk localization of all the eigenstates.

\section{Dice-Haldane Lattice Model}

\begin{figure*} [hbt!]
    \centering
    \includegraphics[width=0.7\textwidth]{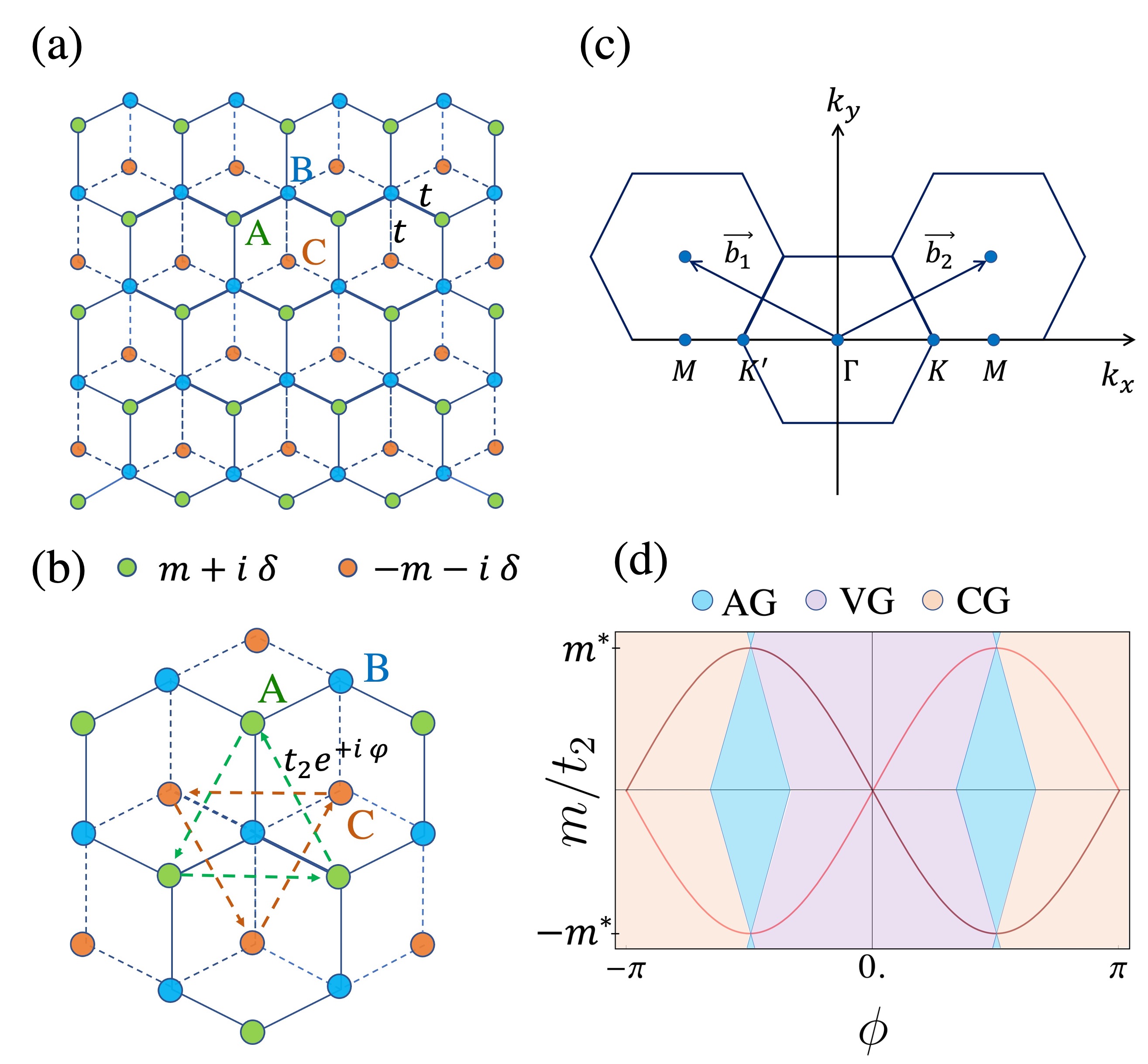}
    \caption{\label{Fig: 1} \textbf{Illustrating the model, its Brillouin zone and the Hermitian phase diagram.} (a) Schematic of the non-Hermitian dice-Haldane lattice model. A, B and C are the three sublattice sites. Hopping potential $t$ is the nearest neighbour hopping between sublattices A-B and B-C. Lattice sites A (shown in green) in general can have a Semenoff mass $+m$ and a non-Hermitian gain $+i \delta$ while lattice sites C (orange) can have a Semenoff mass $-m$ and non-Hermitian loss $-i \delta$. Panel (b) illustrates the next-nearest neighbour Haldane-type hopping with strength $t_2$ and the flux enclosed by these hopping potentials is $\phi$. (c) The BZ of the dice-Haldane lattice showing the high symmetry points $M,K',K$ and $\Gamma$. (d) The phase diagram of the Hermitian dice-Haldane model where the region enclosed within the curves $\pm m^* \sin \phi$ has a non-trivial Chern number. Outside the curve lies the topologically trivial region. Further, we can find three additional phases -- AG: all-gapped (shown in blue), VG: valence-gapped (shown in violet) and CG: conduction-gapped (shown in orange). Here $m^*$ is expressed in units of $t_2$.}
\end{figure*}

In this section, we introduce the Haldane model applied to the dice lattice. The dice lattice can be viewed as a honeycomb lattice with an additional atom at the centre of each hexagonal plaque, as depicted in Fig.~\ref{Fig: 1}(a). Therefore, each unit cell of the dice lattice possesses three basis atoms denoted by A, B, and C in our work. Among these three lattice sites, A and C are equivalent with coordination number three, while B lattice sites have a coordination number of six. There are two main schemes to obtain dice lattice -- using cold atoms confined in optical lattices \cite{bercioux2009massless} and growing a trilayer structure of cubic lattices \textit{viz.} SrTiO$_3$-SrIrO$_3$-SrTiO$_3$ along crystallographic (111) direction \cite{wang2011nearly}. Under the tight-binding framework, this model allows the nearest neighbour (nn) hopping ($t$) between the sites A-B and B-C. Further, in the spirit of the Haldane model, we consider a complex next-nearest neighbour (nnn) hopping ($t_2$) among the A and C lattice sites, such that there is a non-zero flux enclosed by the path formed by the nnn hopping terms. Hence, $t_2 \rightarrow t_2 e^{\pm i \phi}$ where, $\phi$, $+$ and $-$ indicate staggered flux and the sign of the phase for counterclockwise and clockwise hopping about B lattice sites, respectively. Additionally, a Semenoff mass $+m$ on A and $-m$ on C brings us to the full dice-Haldane lattice model. The complete lattice model and the corresponding BZ are schematically represented in Fig.~\ref{Fig: 1}(b) and (c) respectively for convenience (note that for the Hermitian case: $\delta = 0$). The full Hamiltonian of the lattice model can be expressed as follows

\begin{equation}
H= H_{nn} +H_{nnn} +H_m,
\label{eq:hamadd}
\end{equation}

where $H_{nn}$ corresponds to the contribution from only nearest-neighbour hopping. In our calculations, we fix $t=1/\sqrt{2}$, which sets the energy scale of our system. $H_{nnn}$ corresponds to the contribution from next-nearest neighbour hopping parameters and $H_m$ to the asymmetric Semenoff mass terms on A and C sites. The expressions for the same are given by

\begin{equation}
\begin{split}
    H_{nn} &= t \sum_{\langle i,j \rangle} (c_{A,i}^\dagger c_{B,j} + c_{B,i}^\dagger c_{C,j} +h.c.), \\
    H_{nnn} &= t_2 e^{\pm i \phi} \sum_{\langle \langle i,j \rangle \rangle} (c_{A,i}^\dagger c_{A,j} + c_{C,i}^\dagger c_{C,j} +h.c.), \\
    H_m &= m \sum_i (c_{A,i}^\dagger c_{A,i} - c_{C,i}^\dagger c_{C,i}) ,
\end{split}
\label{eq:hameach}
\end{equation}

where, $c_{i}^\dagger$ and $c_{i}$ represent creation and annihilation operator at $i-$th lattice site respectively and $h.c.$ indicates the Hermitian conjugate partner of the given expression. Moreover, $\langle.\rangle$ and $\langle \langle . \rangle \rangle$ denote nearest neighbours and next nearest neighbours, respectively. It is needless to mention that $H_{nnn}$ breaks the time-reversal (TR) symmetry in the model without the requirement of a net external magnetic flux, while $H_m$ breaks the inversion symmetry.

The low-energy electronic description of this dice model can be expressed as a Dirac-Weyl Hamiltonian with pseudospin equal to one. Both the time-reversal symmetry and inversion symmetry broken phases open a gap in the energy spectra. However, these two gapped states are topologically distinct \cite{bandyopadhyay2020topology,bandyopadhyay2020review}, classified by different Chern numbers. In other words, the Hermitian dice-Haldane model \cite{23} harbours a richer phase diagram than the conventional Haldane model, because the former gives rise to more phases both within and outside the topological region of the usual phase diagram accommodating Chern numbers $\pm 2$, as shown in Fig. \ref{Fig: 1}(d). Particularly, the topologically non-trivial region is bounded by the relation $m=\pm m^* \sin \phi$ and has a Chern number $ \pm 2$. Due to the bulk-boundary correspondence, two edge modes will appear when open-boundary conditions are invoked. There are additional phases that arise due to the dice lattice structure and its flat band -- the all-gapped phase (AG), where all three bands are gapped with no overlap, the valence-gapped phase (VG), where the conduction band and flat band touch each other while the valence band remains gapped, and lastly the conduction-gapped phase (CG) where the conduction band is gapped while the valence and flat bands have some overlap. The electronic band structures of the Hermitian dice-Haldane model in the different phases AG, VG and CG can be found in Appendix. \ref{apd1}.

\section{Effect of non-Hermitian Gain and Loss}

Having acquainted ourselves with the Hermitian model, we now systematically invoke non-Hermitian gain and loss into the dice-Haldane model and study its interplay with $H_{nn}$, $H_{nnn}$ and $H_m$. Physically, non-Hermitian balanced gain and loss can be thought of as a source attached to one unit cell atom and an equally strong sink attached to another atom. Here, we allow sublattice A sites to posses a gain $+i \delta$ while sublattice C sites posses a loss $-i \delta$. Hence, the non-Hermiticity added to the Hamiltonian can be described by considering an additional term to Eqn.~\ref{eq:hamadd}, which is of the form

\begin{equation}
        H_\delta= i \delta \sum_i (c_{A,i}^\dagger c_{A,i} - c_{C,i}^\dagger c_{C,i}) .
\end{equation}

Here, $\delta$ denotes the strength of non-Hermiticity. A point to note is that when we turn off $t_2$ and $m$, the Hamiltonian $H$ has parity-time ($PT$) symmetry. Any non-zero value of $t_2$ or $m$ breaks the $PT$ symmetry. 

In this section, we will first consider $H_{nn}+H_{nnn}$ and introduce $\delta$, and thereafter, we will study the effect of non-Hermitian gain and loss considering the full dice-Haldane Hamiltonian $H=H_{nn}+H_{nnn}+H_m+H_\delta$. For each of these cases, we will highlight the exotic physics arising at the high-symmetry points $M, K$ and $\Gamma$, extensively discussing the occurrence of higher-order EPs at integer $\delta$ values. \\

EPs occur when not only two or more eigenvalues become degenerate but also their corresponding eigenvectors \cite{Heiss_2012,Heiss_2004}. This leads to a collapse of the Hilbert space into a lower-dimensional Hilbert space. The collapse of two eigenvectors leads to a second order EP (EP2). A third order EP (EP3) occurs on the collapse of three eigenvectors and so on. The coalescence of eigenvectors can be characterized by the phase rigidity, which is a measure of the bi-orthogonality of the eigenfunctions. It is given by \cite{30,28,49}

\begin{equation}
    r_\alpha= \frac{\langle \phi_\alpha | \psi_\alpha \rangle}{\langle \psi_\alpha | \psi_\alpha \rangle},
\end{equation}

where $\psi_\alpha$ is the $\alpha$-th right eigenvector of $H$ while $\phi_\alpha$ is the corresponding left eigenvector of $H$, i.e $H|\psi_\alpha \rangle = \lambda_\alpha |\psi_\alpha \rangle $ and $\langle \phi_\alpha| H= \lambda_\alpha \langle \phi_\alpha|.$ For a Hermitian system $r_\alpha$ is always equal to unity as the right and left eigenvectors are the same. For non-Hermitian systems, near an EP, $r_\alpha \rightarrow 0$ for the states that coalesce.

Further, to determine the order of the EP, we can perform a scaling analysis of the phase rigidity \cite{71,72}. Here, the Hamiltonian depends on two parameters ($k_x$, $\delta$). The scaling of phase rigidity follows $|r| \sim |\delta- \delta_{EP} |^\nu$ for an $N$-th order EP,  where $\delta_{EP}$ is the value of $\delta$ for which an EP occurs. It is noteworthy that when an anisotropic EP is approached from two orthogonal directions in parameter space the scaling exponent $\nu$ can take values $(N-1)$ or $(N-1)/2$ \cite{xiao2019anisotropic,ding2018experimental}. In our case, we fix $k_x$ and investigate the scaling of the phase rigidity with respect to varying non-Hermiticity $\delta$ close to the EP. Hence, $\nu$ here, is given by $(N-1)/2$ where $N$ is the order of the EP. In particular, we can plot $\log |r|$ vs $\log |\delta-\delta_{EP}|$ to obtain $N$ from the slope. For example, an EP2 will have a slope of 1/2, while an EP3 will have a slope of 1.

\begin{figure*} [hbt!]
    \centering
    \includegraphics[width=0.9\textwidth]{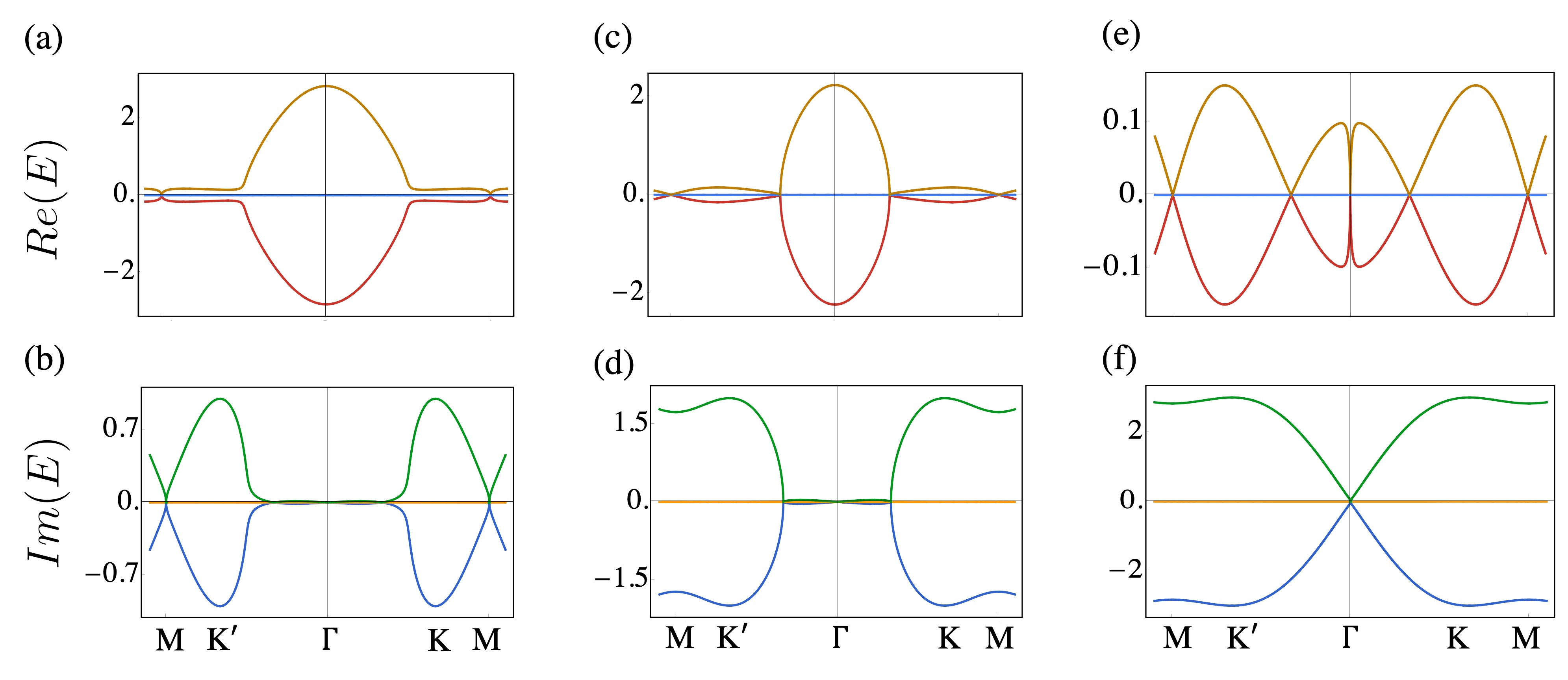}
    \caption{\label{Fig: 6} \textbf{Spectra with next-nearest-neighbour hopping and non-Hermiticity.} The $Re(E)$ (upper panel) and $Im(E)$ spectra  (lower panel) have been shown for different values of non-Hermiticity strength $\delta$. In (a) and (b) $\delta=1.0$ showing an EP at the $M$ point. In (c) and (d) $\delta=2.0$, this induces an EP at $k_x=1/3$. In (e) and (f) we find an EP at the $\Gamma$ point for $\delta=3.0$. For all plots $t=1/\sqrt{2}$, $t_2= 0.06 t$ and $\phi=\pi/2$.}
\end{figure*}

\begin{figure} [hbt!]
    \centering
    \includegraphics[width=0.47\textwidth]{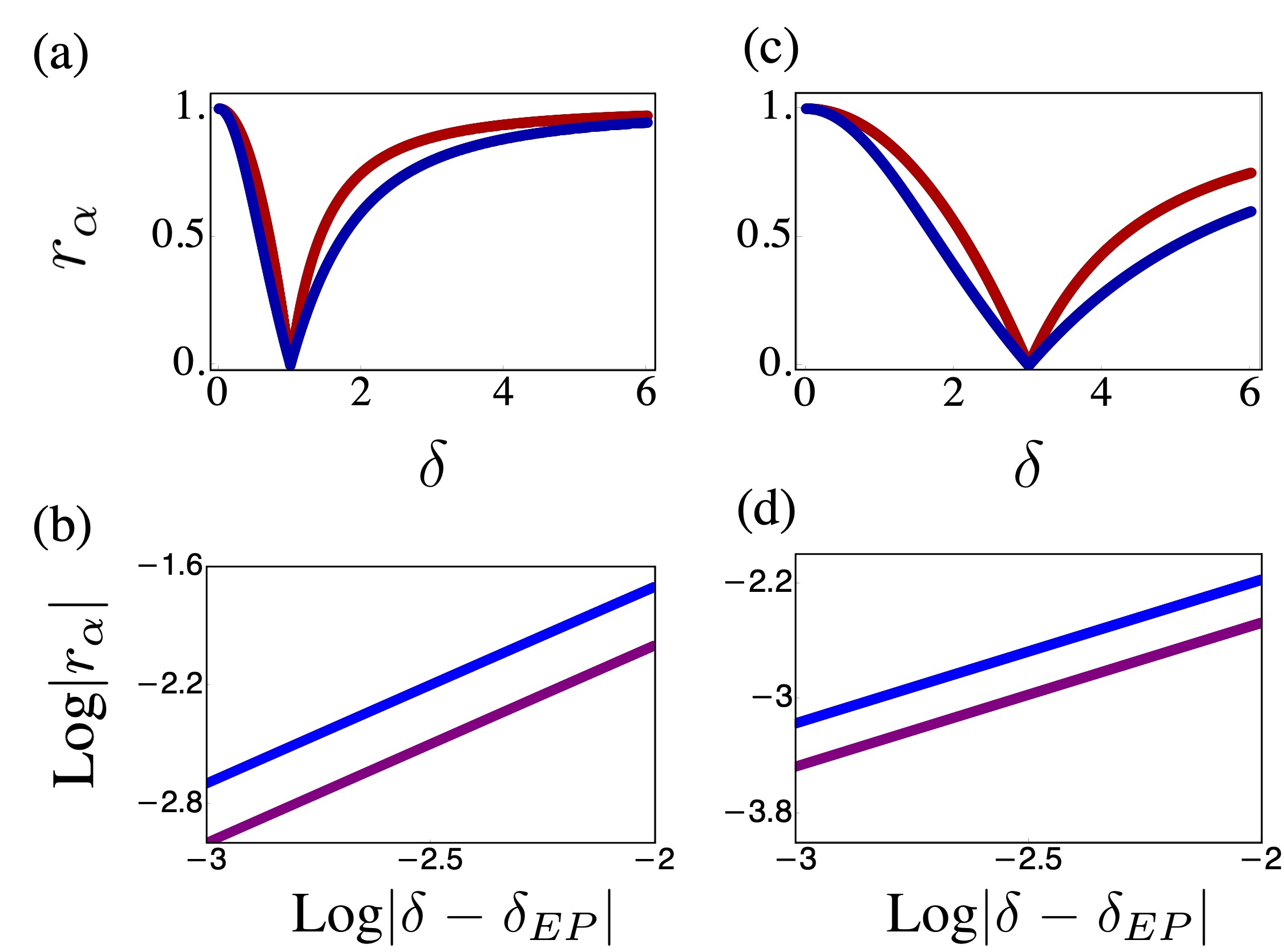}
    \caption{\label{Fig: 7} \textbf{Phase rigidity for next-nearest-neighbour hopping with non-Hermiticity.} The upper panels show the phase rigidity, $r_\alpha$, as a function of non-Hermiticity, $\delta$, while the lower panels show the scaling of the corresponding $r_\alpha$. Panels (a) and (b) correspond to the $M$ point, where an EP is induced at $\delta=1$. Panels (c) and (d) correspond to the $\Gamma$ point, where another EP is induced at $\delta=3.0$. The scaling of $r_\alpha$ gives a slope of one in both cases implying the both the EPs are of order three. The different colours in the plots correspond to different eigenstates. The $r_{\alpha}$ for the two dispersive bands overlap. Here, $t=1/\sqrt{2}$, $t_2= 0.06 t$ and $\phi=\pi/2$.  }
\end{figure}

In addition to the nearest neighbour hopping $t$, first, we include the next-nearest neighbour Haldane type hopping, which breaks time-reversal symmetry of the system. In the presence of a balanced gain and loss term, the Hamiltonian can be written as $H_{nn}+H_{nnn}+H_\delta$. The time-reversal symmetry breaking induces a non-trivial band gap in the Hermitian system (discussed in Appendix \ref{apd1}). In other words, the degeneracy at $K$ and $K^{'}$ points is lifted by non-vanishing $t_2$ (let, $\phi= \pi/2$). In the non-Hermitian case, even a small value of $\delta$ produces a complex energy spectrum as expected. Similar to the nearest neighbour case, finite imaginary parts of the spectra first appear around $K$ and $K^{'}$ points. On the other hand, the conduction and valence bands of the real part of the eigenspectra come closer with increasing $\delta$ and finally meet again at $M$ point for $\delta=1.0$. These observations are illustrated in Fig. \ref{Fig: 6}(a) and (b). Further, the degeneracy of $Re(E)$ at $M$ point is found to be robust to the values of $\delta$, for $ \delta \geq 1$. However, the degeneracy for $Im(E)$ is lifted beyond $ \delta = 1$. With further increase in $\delta$, we have found another set of degeneracies of $Re(E)$ at the halfway point between $\Gamma \rightarrow K /K^{'}$ for $\delta = 2$. Beyond this value of $\delta$, the degeneracy of $Im(E)$ at the same ($\pm1/3,0$) point is removed [Fig. \ref{Fig: 6}(c) and (d)]. It is important to note that the above mentioned degeneracy of $Re(E)$ is robust for $ \delta \geq 2$  similar to what happens at the $M$ point. Finally, at $\delta \geq 3$ degeneracy at $\Gamma$ point appears and disappears for $Re(E)$ and $Im(E)$, respectively, as presented in Fig. \ref{Fig: 6}(e) and (f). Similar to the previous cases, the degeneracy of $Re(E)$ at $\Gamma$ is robust after this critical point. There occur critical values of $\delta$ at which the degeneracy in $Re(E)$ and $Im(E)$ exist simultaneously at particular $k_x$ values. Now, we explore the possibility of these points being EPs and their corresponding order at these critical $\delta$ values, which are interestingly exact integers. The phase rigidity and its corresponding scaling at $M$ and $\Gamma$ points are shown in Fig. \ref{Fig: 7}. From the zero value of $r_{\alpha}$ and the corresponding scaling giving a slope of one, it is clear that all these points are indeed higher-order EPs of order three. 

At this point, it is worth exploring the underlying reason behind the emergence of EPs at high symmetry points due to odd integer values of $\delta$. However, it is possible to get EPs away from high symmetry points at even values of $\delta$. For this purpose, we have analytically calculated the energy band dispersion of the dice lattice described by $H_{nn}+H_{\delta}$. It is evident that the Hamiltonian will give rise to three energy bands including the non-dispersive flat band at zero energy. The dispersive bands, on the other hand, have the following expression.

\begin{equation}
 E_{\pm}(k_x,\delta) = \pm \sqrt{3 - \delta^2 + 4 \cos(\pi k_x) + 2 \cos(2 \pi k_x)}.
\end{equation}

When these bands collapse with the flat band it gives rise to EPs at specific strengths of non-Hermiticity $\delta$ given as below,

\begin{equation}
 \delta_{EP} = \pm \sqrt{3 + 4 \cos(\pi k_x) + 2 \cos(2 \pi k_x)}.
\label{eq:ana1}
\end{equation}

Hence, from the solutions of the above equation we see that EP3 arises at $\delta_{EP} = \pm 3$ and $\delta_{EP} = \pm 1$ for the $\Gamma$ ($k_x = 0$) and $M$ ($k_x = 1$) points respectively. 

Motivated by our above findings, we next add the inversion breaking Semenoff mass term for setting up the complete dice-Haldane lattice model with the Hamiltonian $H=H_{nn}+H_{nnn}+H_m$. We note that the mass term ($+m$ on $A$ lattice sites and $-m$ on $C$ lattice sites) leads to a critical value of $m$ in the units of $t_2$ ($m=m^*=0.16$), where a gap-closing occurs at the $K$ point while the $K'$ point remains gapped. Away from this critical $m$ value, the bands become gapped again. In particular, for $m<m^*$, one lies in the non-trivial topological region of the phase diagram whereas, for $m>m^*$ topologically trivial spectra are obtained. The $m=m^*=0.16$ point corresponds to the semi-metallic phase associated with band gap closing only at $K$ but not at $K'$. Corresponding band diagrams have been detailed in Appendix. \ref{apd1}.

\begin{figure*} [hbt!]
    \centering
    \includegraphics[width=0.9\textwidth]{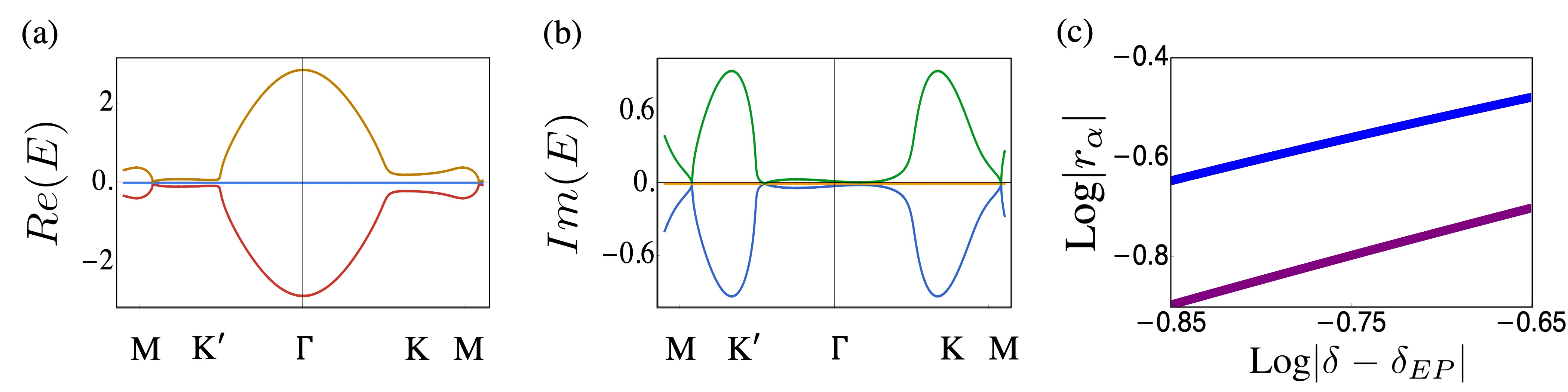}
    \caption{\label{Fig: 9} \textbf{Non-Hermiticity in the complete dice-Haldane lattice model.} In the topologically non-trivial region $m<m^*$, non-Hermiticity strength $\delta$ induces an EP of order three close to the $M$ point in the spectrum. Panel (a) shows the real part of energy spectrum, panel (b) shows the imaginary part of the energy spectrum, and panel (c) shows the scaling of the phase rigidity around $\delta_{EP}=0.94$ at $k_x=1.07$. The different colours here correspond to different eigenstates. The $r_{\alpha}$ for the two dispersive bands overlap. Here, $t=1/\sqrt{2}, t_2=0.06t, \phi=\pi/2$ and $m=0.06$.}
\end{figure*}

\begin{figure}[hbt!]
    \centering
    \includegraphics[width=0.33\textwidth]{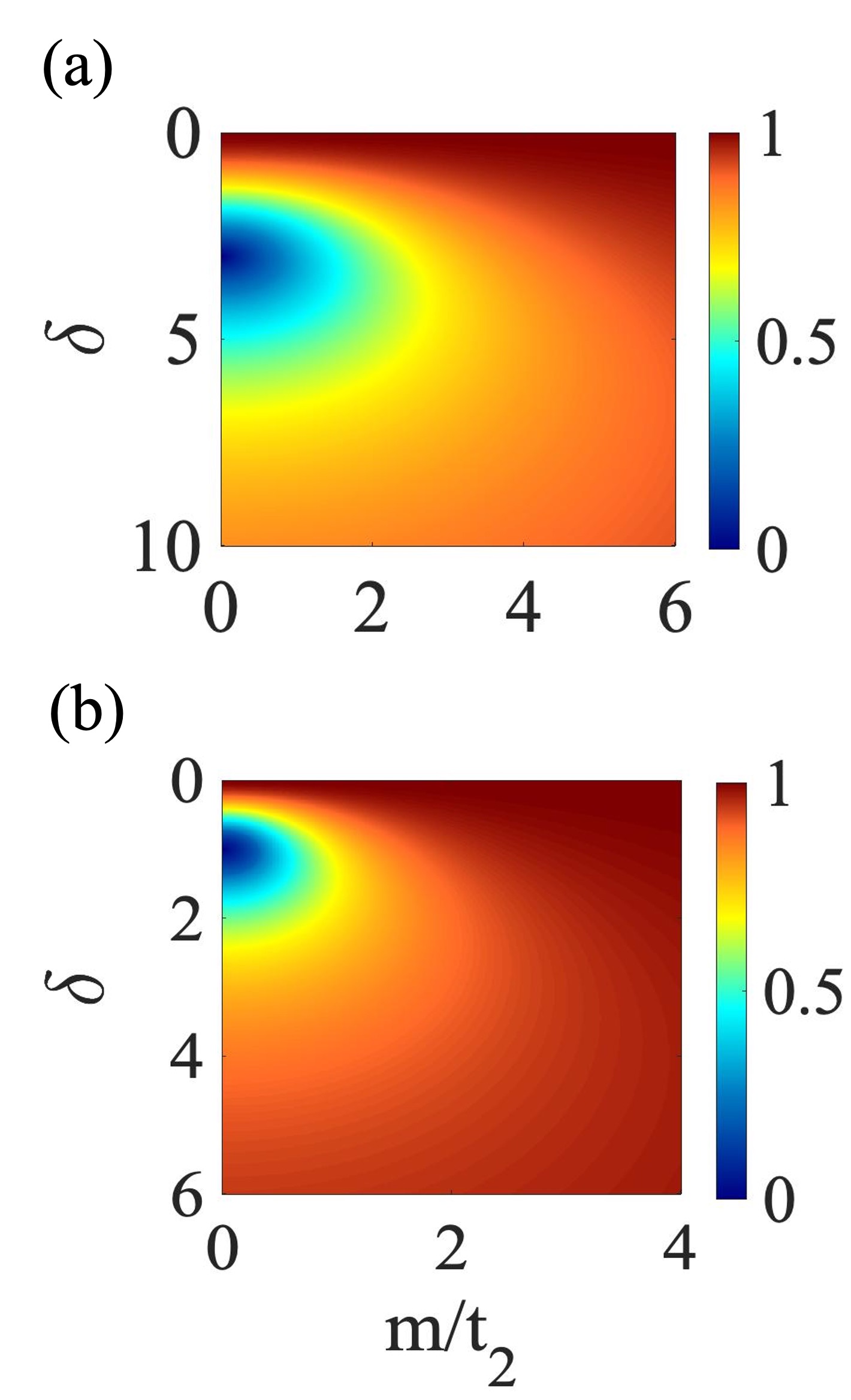}
    \caption{\label{Fig: PhaseRigidity} \textbf{Phase diagram showing the occurrence of EPs as a function of the model parameters.} Panel (a) shows the phase rigidity at the $\Gamma$ point while panel (b) corresponds to the $M$ point. Zero values of phase rigidity imply the existence of third-order EPs, at the corresponding parameter values. Note the extended region in the parameter space with phase rigidity values very close to zero. Here, $t=1/\sqrt{2}$ and $\phi=\pi/2$.}
\end{figure}

We will next explore the effect of the non-Hermitian gain and loss in both the topologically non-trivial and trivial phases. For this purpose, we first chose a value of $m$ ($m = 0.06$) that satisfies the $m<m^*$ criterion for being topologically non-trivial. Further, we introduce and systematically vary $\delta$ to investigate its effect on the complex energy band structure. We find that the sole variation of non-Hermiticity strength $\delta$ can bring about a gap-closed real energy spectrum. This gap closing takes place close to $M$ point for $\delta \sim 1.0$ as presented in Fig. \ref{Fig: 9}(a). Here, the imaginary spectrum is also triply degenerate [Fig. \ref{Fig: 9}(b)], which subsequently gaps out. Hence, we find a third order EP at $k_x=1.07$ for $\delta_{EP}=0.94$. The scaling of the phase rigidity around this EP is shown in Fig. \ref{Fig: 9}(c), confirming its nature. Furthermore, we choose $m$ ($m>m^*$) such that we start from the topologically trivial phase and then invoke non-Hermiticity. In this condition the $Re(E)$ spectra never undergo band closing, and thus EPs cannot emerge, even for arbitrarily large values of $\delta$. Therefore, we have discovered that inversion symmetry breaking in the dice Haldane model offers an EP near the $M$ point only in the topologically non-trivial case. On the other hand, an EP at $\Gamma$ point can be obtained primarily in the inversion-symmetric conditions, i.e., $m=0$. For better understanding, the complete phase diagram for the emergence of EPs at the $\Gamma$ and $M$ points in the parameter space of the dice-Haldane model is presented in Fig.~\ref{Fig: PhaseRigidity}(a) and (b), respectively. The regions in the parameter space where the phase rigidity approaches zero are the regions where EPs can be found. We observe extended regions with phase rigidity values very close to zero. This indicates that fine tuning of parameters $\delta$ and $m/t_2$ is not required to obtain these exceptional regions expanding the possibilities for reaching low values of phase rigidity~\cite{li2022multitude}. The occurrence of EPs also signify a topological phase transition as these are regions of band-gap closing. It is important to note that these phase transitions are possible only at low values of $m$ ($m<m^*$). This follows from our prior observation that gap closings cannot occur solely due to non-Hermiticity unless we are in the topologically non-trivial region of the Hermitian dice-Haldane model. In contrast to the conventional Haldane model, the topological phase transition is now driven by a complex mass term. In particular, the edge states that usually occur in the Hermitian topological phase of Haldane model are also observed in the presence of gain and loss ($\delta$). However, the topological protection of the edge states in the $\mathcal{PT}$ symmetry broken phase holds only up to a critical value of $\delta$. This phase transition is associated with passing through a third-order EP. On the other hand, if we choose the value of mass $m$ outside the topologically non-trivial region of the Hermitian case, there is no possible value of $\delta$ that will manifest in protected edge states and will trace the system back into the topological region. In other words, to obtain EPs at any finite, non-zero value of $\delta$, we require the Hermitian system to be placed initially within the topological region. Tuning $\delta$ can bring about a topological phase transition enabling the occurrence of EPs in the non-Hermitian model.

Now, we will invoke a different class of non-Hermiticity -- non-reciprocal hopping along one direction. In particular, we have assigned $t+\gamma$ ($t-\gamma$) to the hopping parameters $C \rightarrow B$ ($B \rightarrow C$) and $B \rightarrow A$ ($A \rightarrow B$) in the vertical direction in Fig. \ref{Fig: 1}(a). Even in the absence of the non-Hermitian gain and loss we observe that non-reciprocal hopping solely can induce EPs in the system at specific strengths of non-Hermiticity $\gamma$. The deformation of the electronic bands under this non-reciprocity is qualitatively similar to our previous results with non-Hermitian gain and loss. However, the critical values of $\gamma$ at which the EPs occur differ from the critical $\delta$ we found in the above discussion. For example, in the case with only nearest neighbour hopping EPs are induced at critical $\gamma$ values of

\begin{equation}
    \gamma_{EP} = \pm \sqrt{\frac{3}{2} + 2 \cos(\pi k_x) + \cos(2 \pi k_x)}.
\end{equation}

Therefore, from Eqn.~\ref{eq:ana1} it is clear that $\gamma_{EP}$ values are related to the corresponding $\delta_{EP}$ values by the relation $\gamma_{EP} = \delta_{EP} /\sqrt{2}$. Numerically, we have verified the occurrence of EP3 at $\gamma_{EP} = \pm \frac{3}{\sqrt{2}}$ and $\gamma_{EP} = \pm \frac{1}{\sqrt{2}}$ for the $\Gamma$ and $M$ points respectively. It is interesting to study this non-reciprocal hopping in the context of the dice-Haldane nanoribbon to see the effects of OBC [see Section \ref{sec: nano}(B)].

\section{Finite-size effects: Dice-Haldane nanoribbon \label{sec: nano}}

\begin{figure} [hbt!]
    \centering
    \includegraphics[width=0.45\textwidth]{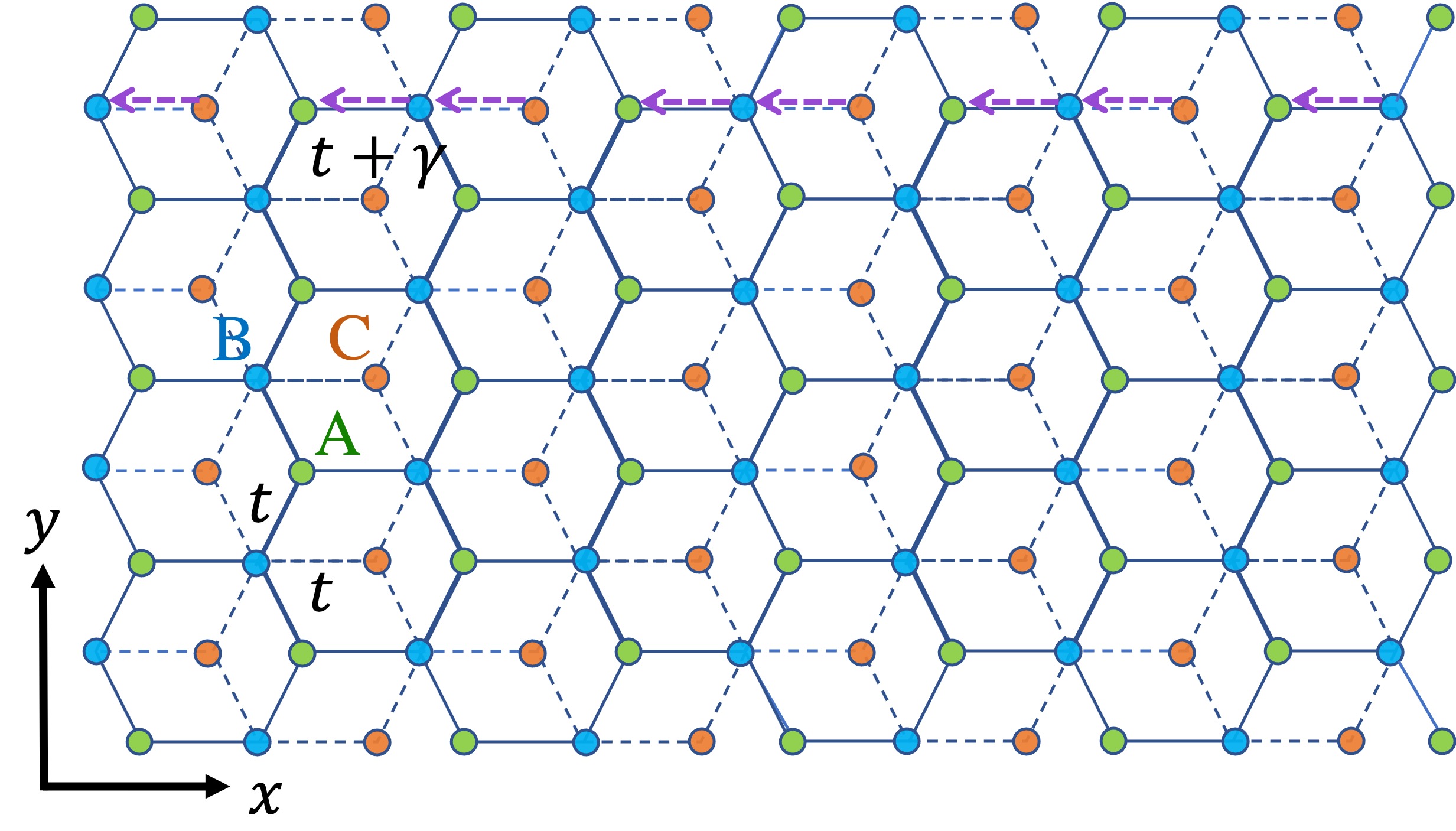}
    \caption{\label{Fig: 10} \textbf{Schematic of the dice-Haldane nanoribbon with non-reciprocal hopping.} The hopping and on-site parameters remain the same as in the dice-Haldane sheet case. An additional kind of non-Hermiticity $\gamma$ has been introduced in the nanoribbon, namely a non-reciprocal hopping. This favours nearest-neighbour hopping from right to left along the $x$ direction rather than from left to right.}
\end{figure}

Having understood the effect of non-Hermiticity in the $k$-space model of the dice lattice sheet, we next move on to the study of another physically important case of the dice-Haldane nanoribbon extended along one direction (say $x$). The schematic of the finite size nanoribbon considered here is shown in Fig. \ref{Fig: 10}. Here $L_x$ and $L_y$ are the dimensions of the nanoribbon in the $x$ and $y$ directions, respectively. We particularlt focus on a real space model of the nanoribbon, where both $L_x$ and $L_y$ are finite. The real space model with open-boundary conditions in both directions leads to interesting consequences when we invoke non-Hermiticity as we will discuss shortly.

Considering the Hermitian dice-Haldane model, we can expect that in the topologically non-trivial phase, each edge of the lattice will exhibit two chiral edge states since the Chern number is $\pm 2$. These edge modes lie in the band gap and connect two bulk bands. In contrast to the honeycomb lattice, the spectrum exhibits two unidirectional chiral states per edge for $m<m^*$ that cross over from the bulk states near the Fermi level. However, for $m>m^*$, the edge states near the flat band are counter-propagating at a given edge. Hence, no net current will flow through it. Consequently, the bulk states remain gapped out, and the bulk boundary correspondence continues to hold. Corresponding figures and a discussion can be found in Appendix. \ref{apd3}.

It is important to note that these edge states are quite robust to both real and complex on-site disorder, i.e., despite some disorder-induced distortion in the shape of the bands, the edge states persist up to large values of disorder. We have checked this for a disorder of the form $\Delta_j$ on each lattice site, where $\Delta_j$ is allowed to be real or imaginary, corresponding to real and imaginary on-site disorder, respectively. Here, $j$ denotes the lattice site and $\Delta_j= \Delta \omega_j$ where $\omega_j \in [-1,1]$. The disordered Hamiltonian has the form $H=H_{nn}+H_{nnn}+H_m+H_{dis}$ where,

\begin{equation}
    H_{dis}= \sum_j \Delta_j c_j^\dagger c_j. \label{Eq:dis}
\end{equation}

We will study the effect of this complex random on-site disorder in more detail when we introduce non-Hermiticity in the finite nanoribbon geometry. We address the interplay of non-Hermiticity and disorder in Section. \ref{sec ribbon gamma}.

\subsection{Nanoribbon with non-Hermitian gain and loss}

We first consider the effect of non-Hermitian balanced gain and loss in the dice-Haldane nanoribbon. The Hamiltonian under such considerations is given by $H=H_{nn}+H_{nnn}+H_m+H_\delta$. It is worth noting that the topological edge states found in the Hermitian regime, forming conducting channels between the conduction and valence bands are robust even in the presence of non-Hermiticity. For a range of increasing values of $\delta$, up to a system dependent critical value $\delta_c$, the topological edge states can be clearly discerned from the energy band diagram. A detailed discussion can be found in Appendix. \ref{apd3}.\\

\begin{figure} [hbt!]
    \centering
    \includegraphics[width=0.5\textwidth]{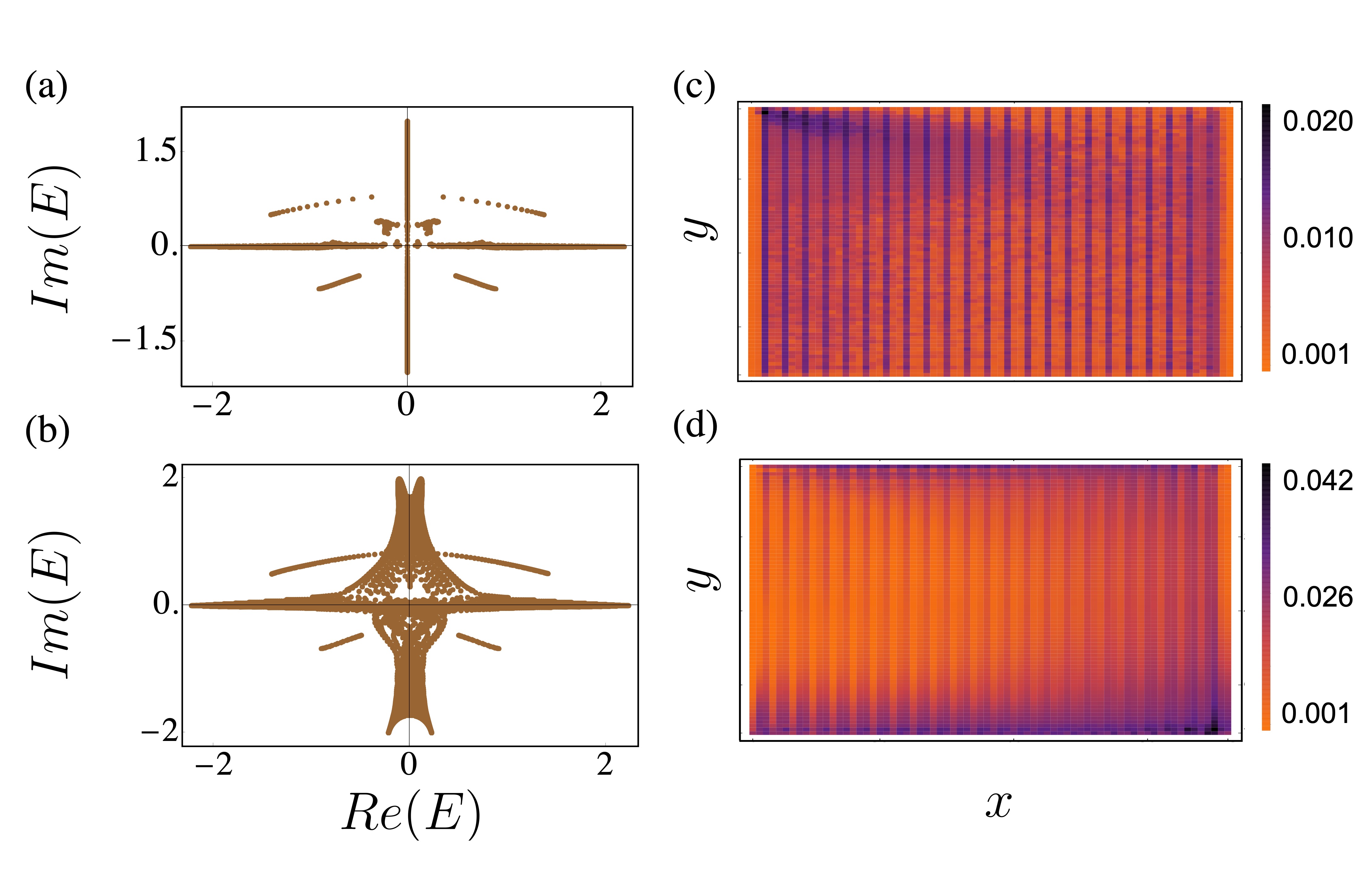}
    \caption{\label{Fig: GL} \textbf{Spectra and LDOS with non-Hermitian balanced gain and loss.} Panels (a) and (b) show the complex energy spectrum for the torus (PBC) geometry of the nanoribbon and panels (c) and (d) show the respective LDOS for the same system under OBC. Panels (a) and (c) correspond to the dice lattice ($t_2=0$) with non-Hermitian balanced gain and loss. The energy spectrum in (a) does not enclose a finite, non-zero area. This implies that under OBC there will be no occurrence of NHSE. The corresponding LDOS of the system under OBC shown in (c) demonstrates the absence of skin effect. Panels (b) and (d) correspond to the full dice-Haldane lattice with gain and loss. Here, $t_2=0.06 t$. The spectrum shown in (b) encloses a finite non-zero spectral area indicating the possibility of an NHSE under OBC, which is established by the LDOS shown in (d). In this case, the localization of states occur at both the top and bottom edges of the nanoribbon. Here, $t=1/\sqrt{2}, \phi=\pi/2$, $m=0$ and gain and loss strength $\delta=2.0$. }
\end{figure}

One of the striking features of non-Hermitian systems has been the discovery of NHSE and it is interesting to understand whether our proposed system exhibits this feature. It may be noted that these fascinating phenomena unique to non-Hermitian systems, such as non-Bloch EPs and skin effects, can be well explained in terms of the generalized Brillouin zone (GBZ) formalism~\cite{51,yang2020non}. We note that it has been established that a two-dimensional system under OBC can exhibit NHSE if and only if under PBC the complex eigenspectrum encloses a finite non-zero spectral area \cite{Zhang_2022}. We consider a periodic version of the system, i.e., a dice-Haldane torus and invoke balanced gain and loss. When we consider the dice lattice with only nearest neighbour interactions under PBC, the complex eigenspectrum of the system has an arc-like structure and does not enclose any finite spectral area in the complex plane, as shown in Fig. \ref{Fig: GL}(a). This indicates the absence of an NHSE when we invoke OBC. In order to probe this, we calculate the local density of states (LDOS) for the system under OBC, i.e., the finite nanoribbon. To calculate the LDOS, we evaluate $\sum_\alpha |\psi_\alpha(x_i)|^2$ at each lattice site ($x_i$) which gives us LDOS($x_i$). The plot of the corresponding LDOS for the dice lattice in Fig.~\ref{Fig: GL}(c) verifies that NHSE is indeed absent and the states are distributed over the lattice. However, when we consider the full dice-Haldane periodic system ($t_2 \neq 0$) with balanced gain and loss, the complex spectrum does enclose a finite area [shown in Fig.~\ref{Fig: GL}(b)]. This translates to the occurence of NHSE when OBC is invoked. Remarkably, the localization of the eigenstates occurs at both the top and bottom edges of the nanoribbon which can be seen fom the plot of the LDOS in Fig.\ref{Fig: GL}(d). So, non-Hermitian gain and loss is able to cause an NHSE in the dice-Haldane model and not in the dice model which accounts for only nearest neighbour hopping. The nature of the skin effect occurring only at the top and bottom edges of the dice-Haldane system [as shown in Fig. \ref{Fig: GL}(d)], can be explained further through the winding number and  the complex energy spectra by imposing OBC in one direction while retaining PBC in the other. A detailed discussion and corresponding figures can be found in Appendix. \ref{apd4}.
For all cases of our computations the dice-Haldane nanoribbon has 72 $\times$ 36 sites ($n=2592$), unless stated otherwise.

\subsection{Nanoribbon with non-reciprocal hopping \label{sec ribbon gamma}}

\begin{figure} [hbt!]
    \centering
    \includegraphics[width=0.5\textwidth]{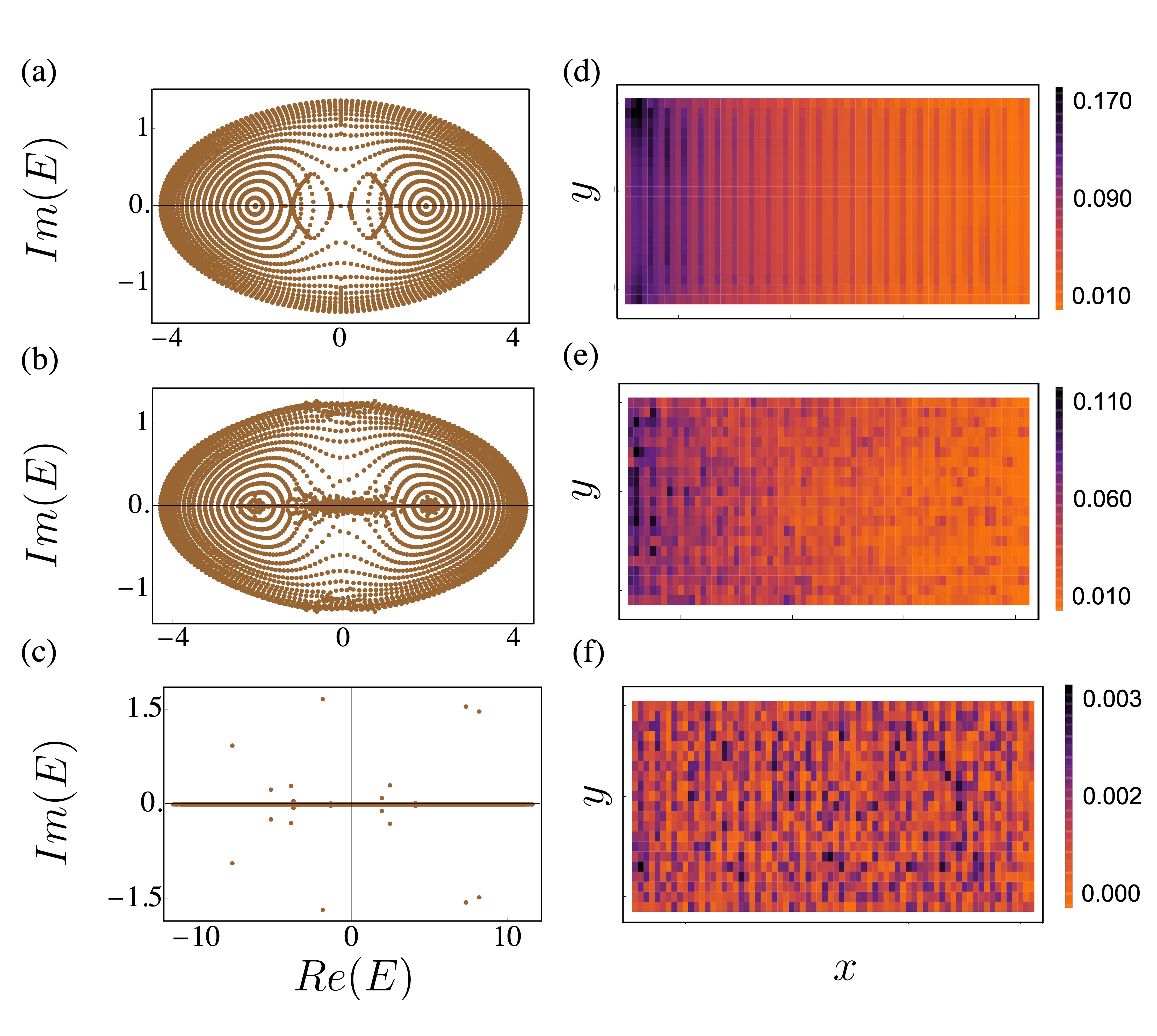}
    \caption{\label{Fig: NRH} \textbf{Effect of disorder on the spectra and LDOS for the dice-Haldane lattice with non-reciprocal hopping.} Panels (a), (b) and (c) show the complex spectra of the dice-Haldane model under PBC with non-reciprocal hopping. Panels (d), (e) and (f) show the corresponding LDOS for the same system under OBC. Panels (a) and (d) show the disorder-free system. Here, the complex eigenspectrum [shown in (a)] encloses a non-zero spectral area which translates to a skin effect under OBC. The corresponding LDOS [shown in (d)] depicts an NHSE, with states accumulating close to $x=1$, which gradually decreases as we move rightward. Panels (b) and (e) correspond to disorder strength $\Delta=1$. Here, the spectral area [shown in (b)] is still non-zero and finite, implying an existence of NHSE, which is established by the LDOS in (e). For panels (c) and (f) where $\Delta=10$, the spectral area has disappeared indicating the destruction of NHSE. The corresponding LDOS shown in (f) attributes to the same where the skin effect has completely disappeared due to localization in the bulk. Here, $t=1/\sqrt{2},t_2=0.06 t, \phi=\pi/2$, $m=0$ and non-reciprocity strength $\gamma=2.0$.}
\end{figure}

As we analyzed previously for the periodic dice-Haldane sheet, in this section, we consider the dice-Haldane nanoribbon and study the effect of non-reciprocal hopping. In particular, non-reciprocal nearest-neighbour hopping is introduced only along the $x$ direction. We have the hopping values $t-\gamma$ along $+x$ and $t+\gamma$ along $-x$ directions, i.e., we have a biased hopping strength that favours hopping from right to left rather than from left to right. We invoke this non-reciprocal hopping throughout the bulk of the nanoribbon [See Fig. \ref{Fig: 10}]. We study the spectrum and LDOS of the system with this type of non-Hermiticity and find strikingly different behaviour than in the case of balanced gain and loss. Unlike in the prior case, when we introduce a non-reciprocal hopping, i.e., $\gamma \neq 0$, the PBC spectrum covers a finite area in the complex plane even in case of the dice model with $t_2=0$, which implies an NHSE under OBC. In fact, the effect of non-reciprocal hopping on the dice model is qualitatively the same as its effect on the dice-Haldane lattice. We will study the latter in detail and also discuss the effect of disorder for this case. 
The spectrum of the dice-Haldane model under PBC with non-reciprocal hopping accommodates a finite spectral area in the complex plane shown in Fig. \ref{Fig: NRH}(a). This indicates the possibility of an NHSE when we impose OBC in the system. Next, we investigate the effect of disorder on this spectral area where the nature of disorder has been taken according to Eqn.~\ref{Eq:dis}. When the disorder strength is taken to be $\Delta=1$, the complex spectrum undergoes some disortion but still accommodates a finite area [Fig.~\ref{Fig: NRH}(b)]. However, at a large disorder strength ($\Delta=10$) shown in Fig.~\ref{Fig: NRH}(c) the spectral area disappears. This indicates that under OBC, the NHSE will gradually get destroyed due to the introduction of disorder. Next, we investigate the behaviour of the LDOS to verify the above findings and to visualise the occurrence and subsequent disappearance of the NHSE under disorder.

In Fig. \ref{Fig: NRH}(d) the LDOS can be seen to be higher around small values of $x$ and decreases as we go to higher $x$. This implies a maximal concentration of eigenstates near the left edge of the system. Thus, non-reciprocal hopping, when invoked throughout the bulk of the nanoribbon, causes an NHSE. It is important to note that the NHSE in this case is different from that caused by gain and loss. Here, the localization of the eigenstates is at one edge (left) of the lattice and is also directionally different than in the previous case where NHSE occurred at the top and bottom edges. In the former case of gain and loss, the inclusion of the Haldane next-nearest neighbour hopping is essential for realizing the skin effect. Here, the staggered magnetic flux in the presence of non-Hermitian gain and loss introduces chiral edge currents in two different directions for the two distinct sublattices A and C \cite{cai2019experimental}. Consequently, the eigenmodes are localized at the top and bottom edges of the ribbon under OBC. In contrast, the non-reciprocal hopping along one direction offers a directionally-biased propagation of eigenstates, causing localization at the left edge. We find that the LDOS does not vary continuously from high to low as we move along $+x$ but shows regions of high value followed by those of lower value. This feature is due to the missing hopping terms between sublattices A and C, which inhibits the complete flow of the eigenstates leftwards. This can be pictured from the schematic in Fig. \ref{Fig: 10}, which suggests that the accumulation of states will be greater on sublattice C and gradually decrease towards the following A lattice site. Now, we look at the effect of disorder on the NHSE. Upon increasing the value of disorder strength, there occurs a localization of the eigenstates, as shown in Fig. \ref{Fig: NRH}(e and f). In Fig.~\ref{Fig: NRH}(e) corresponding to $\Delta=1$ the concentration of eigenstates at the left edge has decreased, signifying a partial destruction of the NHSE. Finally, at large values of disorder ($\Delta=10$), a complete localization of the eigenfunctions is found, causing a low value of the LDOS over all $x$, implying the complete destruction of the NHSE [Fig. \ref{Fig: NRH} (f)]. \\

\begin{figure*} [hbt!]
    \centering
    \includegraphics[width=0.95\textwidth]{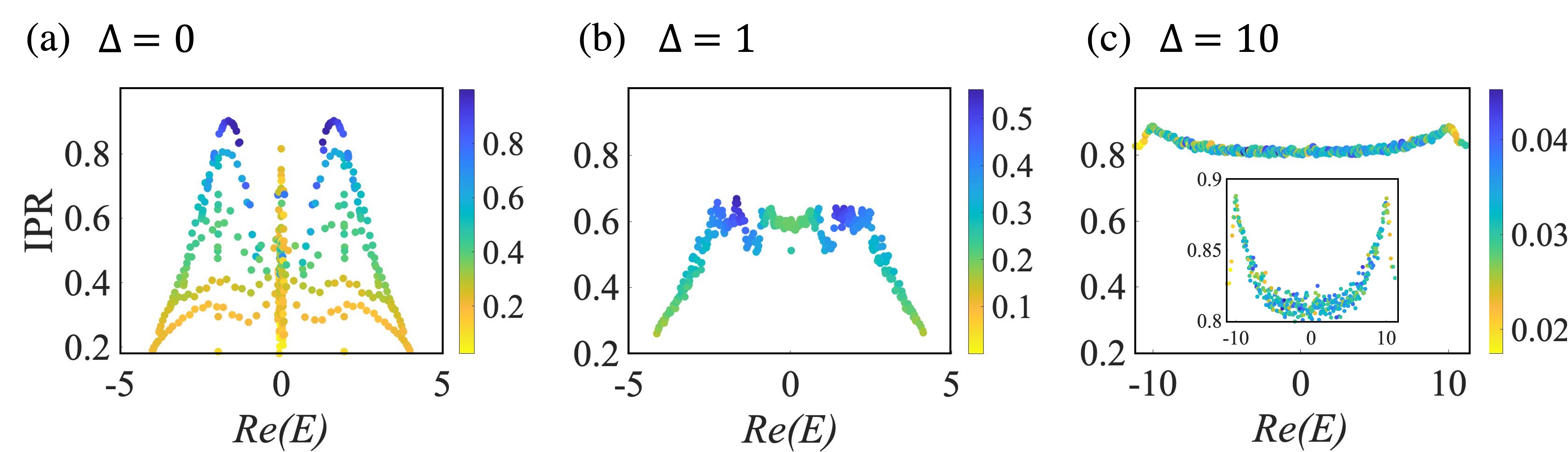}
    \caption{\label{Fig: IPR} \textbf{IPR and edge probability with non-reciprocal bulk.} The IPR has been plotted for all eigenstates as a function of $Re(E)$ and the corresponding colour denotes its probability density at the left edge of the nanoribbon. Panel (a) corresponds to $\Delta=0$, where the IPR is very high for the states localized at the edge and hence establishes the occurrence of NHSE. In panel (b) the IPR values have diminished, implying the reduction of skin effect due to disorder. Here, disorder strength $\Delta=1$. Panel (c) corresponding to $\Delta=10$, shows very high values of IPR but very low edge population implying the occurrence of disorder-induced bulk localization and complete destruction of NHSE. A zoomed in plot is shown in the inset. Here, the IPR and edge probability have been disorder averaged over 1000 configurations. The number of hexagonal layers in the $y$ direction has been taken to be 35. We have set $t=1/\sqrt{2},t_2=0.06 t, \phi=\pi/2, m=0, \delta=0, \gamma=2$.}
\end{figure*}

To consolidate the above arguments, we next study the behaviour of the IPR \cite{kramer1993localization} and the probability density of the eigenstates at the edge after averaging both the quantities over multiple disorder configurations. We averaged over 1000 disorder configurations. The IPR for the $\alpha$-th eigenstate, $I_{\alpha}$, is defined as 

\begin{equation}
    I_{\alpha} = \frac{\sum_{r} \abs{\psi_{\alpha}(r)}^4}{\left(\sum_{r} \abs{\psi_{\alpha}(r)}^2\right)^2}.
\end{equation}

For localized states, IPR is close to 1 while for extended states IPR is very low. We further define the edge probability, $P_{\alpha}$, of state $\psi_{\alpha}$ as 

\begin{equation}
    P_{\alpha} = \frac{\sum_{x_i=1}^{x_E} \abs{\psi_{\alpha}(x_i)}^2}{\sum_{x_i} \abs{\psi_{\alpha}(x_i)}^2},
\end{equation}

where, $x_E$ represents the width of the edge, which we take as the first five lattice sites from the left end of the ribbon ($x_E = 5$). In the case of the Hermitian dice-Haldane nanoribbon the dispersive bands are delocalized with very low IPR and there is no skin effect as expected. In presence of non-reciprocity in the bulk ($\gamma = 2$) we observe IPR $\rightarrow$ 1 for the eigenstates localized at the edge. This can be seen from Fig.~\ref{Fig: IPR}(a) where the edge probability is denoted by the colour bar showing that the states with high $P_{\alpha}$ correspond to high IPR. Hence, for the disorder-free case with non-reciprocal bulk we can confirm the occurrence of NHSE. Next, we look at the effect of disorder on the skin effect -- the presence of disorder essentially reduces the edge localization in the system. This can be discerned from the diminished values of the IPR in Fig.~\ref{Fig: IPR}(b), presented for disorder strength $\Delta = 1$. Yet, the higher values of IPR are predominantly contributed by the eigenstates near the edge. Further, for large disorder strength ($\Delta = 10$), shown in Fig.~\ref{Fig: IPR}(c), the IPR is uniformly high, although $P_{\alpha}$ is very low for all the eigenstates. This corresponds to the disorder induced bulk localization and hence the complete destruction of NHSE. 

Finally, we also note that when we invoke non-reciprocal hopping only along the upper and lower edges of the system and not in the bulk, surprisingly, here too there is the occurrence of the NHSE which is similarly destroyed at large values of disorder.

\section{Summary and discussion}

In this work, we have systematically studied the effect of non-Hermiticity in the Chern insulating dice-Haldane lattice. 

We introduced non-Hermiticity in this model in two ways: (i) using balanced gain and loss terms and (ii) setting non-reciprocal hopping parameters. Introducing non-Hermiticity through any of the above means invariably causes higher-order exceptional points. Our analytical description revealed that the exceptional points at high symmetry points emerge at odd integer values of the gain and loss non-Hermiticity strength and at $1/\sqrt{2}$ times the previous values in the case of non-reciprocal hopping. Further, we showed that the dice-Haldane lattice consisting of complex next-nearest neighbour hopping and the Semenoff mass offers a rich topological phase diagram. The robustness of the topological edge states was critically examined with non-Hermiticity and complex disorder.
Moreover, we discover that, unlike the gain and loss case, the non-reciprocal hopping triggers a fascinating non-Hermitian skin effect under OBC for the dice lattice with only nearest neighbour couplings. However, both kinds of non-Hermiticity can cause NHSE in the more general dice-Haldane nanoribbon. Remarkably, the NHSE caused by gain and loss generates localization at the top and bottom edges while non-reciprocity results in an NHSE at the left edge of the nanoribbon. The directionality of localization of maximal eigenstates can hence be tuned using the nature of non-Hermiticity and its strength. The skin effect is protected by a finite spectral area in the complex plane under PBC in real space. Furthermore, the LDOS, IPR, and edge probability calculations also demonstrate the occurrence of the skin effect and its robustness to the disorder.

Our study is fundamental to understanding the tunability of the dice-Haldane model under the influence of non-Hermiticity, especially in the context of EPs which have been experimentally realized in microwave cavity resonators \cite{PhysRevLett.86.787} and coupled electronic circuits \cite{Stehmann_2004}. Specifically, EPs of order three have been realized in coupled acoustic cavity resonators \cite{PhysRevX.6.021007} and optical cavity systems \cite{hodaei2017enhanced}. It would be interesting to engineer already fabricated dice lattices to introduce non-Hermitian gain and loss or non-reciprocal hopping to obtain these higher-order EPs. \\

The key ingredients for attaining a dice lattice in cold atomic systems are three pairs of counterpropagating lasers placed at an angle of $120^{\circ}$ with respect to each other \cite{bercioux2009massless,moller2012correlated}. This laser setup essentially divides the 2D plane into six equivalent parts. Further, the interference causes standing waves that give rise to the required potential traps of the optical lattice. In particular, a dice lattice with lattice constant $a_0$ can be constructed by using six linearly polarized laser beams of wavelength $\lambda = 3 a_0 /2$. Another plausible pathway for fabricating dice lattices is the use of coupled resonators \cite{zhang2020compact}. The prescription is the incorporation of additional resonators at the center of hexagonal rings of the honeycomb lattice. The ring-shaped primary resonators of the dice lattice are effectively connected with each other via auxiliary resonators placed in between. The Haldane model has been experimentally realized in optical lattices using ultracold atoms \cite{jotzu2014experimental}. 
Notably, the time-reversal symmetry can be broken through complex next-nearest-neighbor tunnelling induced by circular modulation of the lattice position in time. Additionally, the deformation of lattice geometry by applying unidirectional in-plane force with the help of a magnetic field gradient provides an energy offset and breaks the inversion symmetry. 

Non-Hermiticity has been successfully engineered into several optical lattices, acoustic systems and topoelectrical circuits \cite{ ding2016emergence,ding2018experimental,zhang2020non,PhysRevResearch.3.023056}.  In particular, the inclusion of non-Hermiticity in the three-site Lieb lattice using coupled optical waveguides \cite{xiao2020exceptional} can be feasibly extended to our lattice system. It has been established that optical lattices fabricated using femtosecond-direct-laser-writing can adduce non-Hermitian gain and loss through periodic `breaks' in the waveguides, which lead to loss of radiation modes. This loss can be tuned using the length of the breaks.\cite{cerjan2019experimental}. This method of engineering gain and loss has also been successfully realized in a graphene-like honeycomb lattice \cite{rechtsman2013photonic}. Further, it has been proposed that atomic loss in ultracold atomic gas systems can be generated using a resonant optical beam to kick the weakly trapped atoms or by using a radio frequency to excite the atom to an irrelevant state, thereby simulating loss\cite{xu2017weyl}.
On-site gain and loss can be effectively mapped onto a non-Hermiticity-controlled coupling between neighbouring atoms. A synthetic imaginary gauge field engineered strategically can make these couplings asymmetric \cite{longhi2015non}. Such complex gauge potentials causing non-reciprocal hopping can be implemented using a non-Hermitian anti-resonance ring \cite{midya2018non}. Due to the directional coupling, the photons become attenuated or amplified depending on their direction of travel.
Furthermore, two-dimensional non-Hermitian systems with gain and loss or non-reciprocity have been proposed in classical topoelectrical circuits \cite{zhang2020non,PhysRevResearch.3.023056}, where the non-Hermiticity can be ingeniously engineered using combinations of resistances and LC-tanks. Information about the eigenenergies can be extracted from the electrical response, admittance and impedance resonances \cite{lee2018topolectrical}. In light of the above rapid experimental advances, we believe our theoretical findings can be experimentally tested in the near future.

\section*{ACKNOWLEDGEMENTS} R.S. acknowledges Indian Institute of Science for financial support. A.B. thanks the IoE postdoctoral fellowship for support. R.S. and A.B. would also like to thank Ayan Banerjee for several useful discussions. A.N. acknowledges support from the start-up grant (SG/MHRD-19-0001) of the Indian Institute of Science and DST-SERB (project number SRG/2020/000153).


\appendix

\section{Hermitian dice-Haldane band diagrams}\label{apd1}

\begin{figure*} [!hbt]
    \centering
    \includegraphics[width=0.9\textwidth]{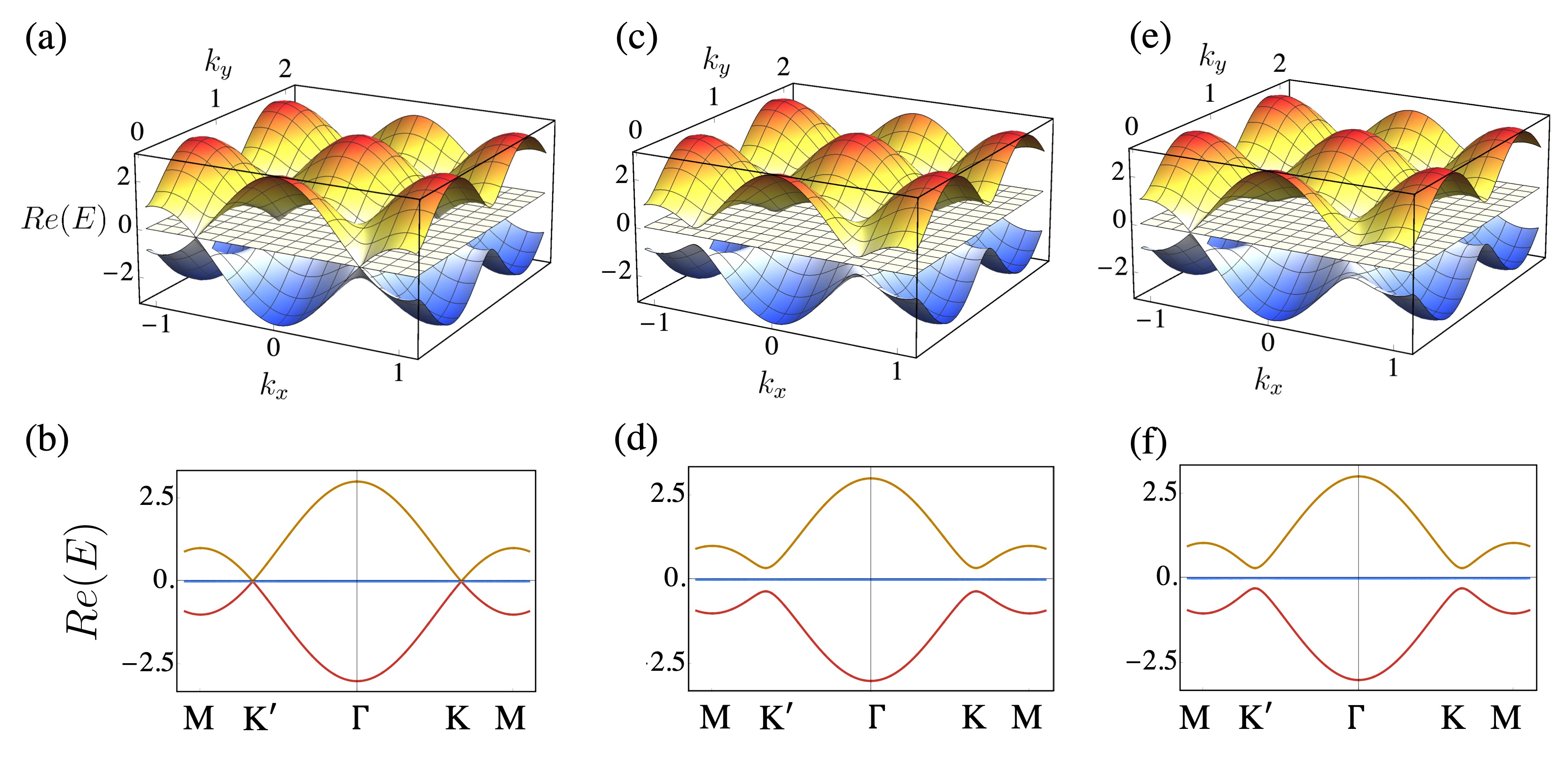}
    \caption{\label{Fig: 3} \textbf{Energy spectra of the Hermitian model.} The band structure of the Hermitian dice-Haldane model along high symmetry points $M-K'-\Gamma-K-M$. The upper panels show the three-dimensional spectrum as a function of $k_x$ and $k_y$, while the lower panels show the corresponding two-dimensional plots at $k_y=0$. Panels (a) and (b) show the effect of only nearest-neighbour hopping with $t_2=m=0$. Here, all three bands are gapless at the $K$ and $K'$ points. In (c) and (d) next-nearest neighbour hopping has been included. Here, $t_2=0.1, \phi=\pi/2$ while $m=0$. In (e) and (f) $m=0.3$ while $t_2=0$ shows the effect of the Semenoff mass term. Both $t_2$ and $m$ open up a gap in the spectrum. For all plots the value of nearest neighbour hopping is chosen to be $t=1/\sqrt{2}$.}
\end{figure*}

\begin{figure*} [hbt!]
    \centering
    \includegraphics[width=0.9\textwidth]{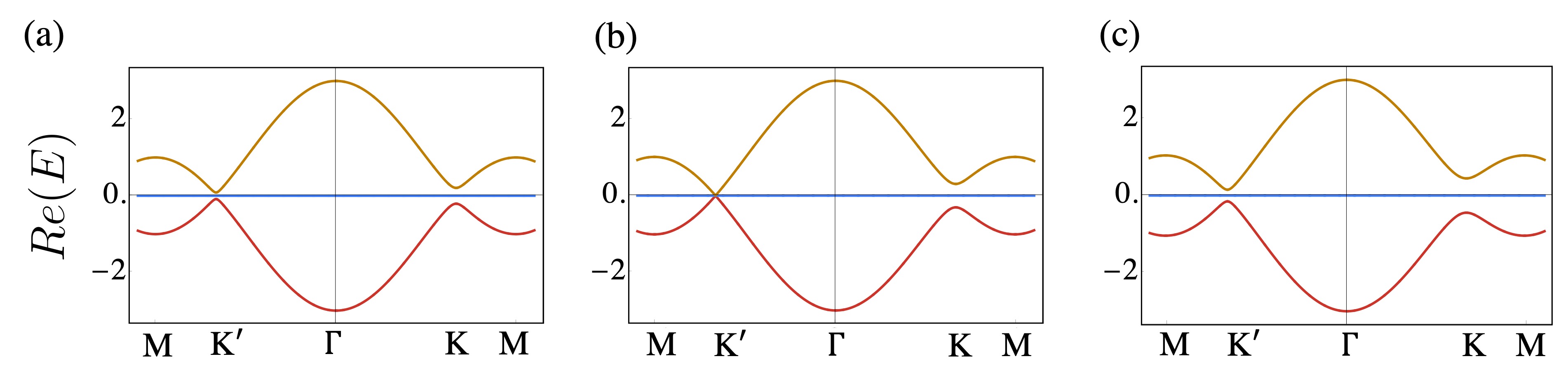}
    \caption{\label{Fig: 8} \textbf{Effect of the Semenoff mass in the Hermitian model with both $t$ and $t_2.$} In (a) $m=0.06$, in (b) $m=0.16$ and in (c) $m=0.30$. An arbitrarily small Semenoff mass opens up a gap in the spectrum which subsequently closes in (b) indicating a semi-metallic phase. On increasing the value of $m$ further, all bands become gapped again. This critical value of $m$ is $m^*=0.16$ which separates the topologically non-trivial and trivial regions (a) and (c), respectively. For all plots $t=1/\sqrt{2}, t_2 = 0.06 t$ and $\phi=\pi/2$. }
\end{figure*}

\begin{figure*}[hbt!]
    \centering
    \includegraphics[width=0.9\textwidth]{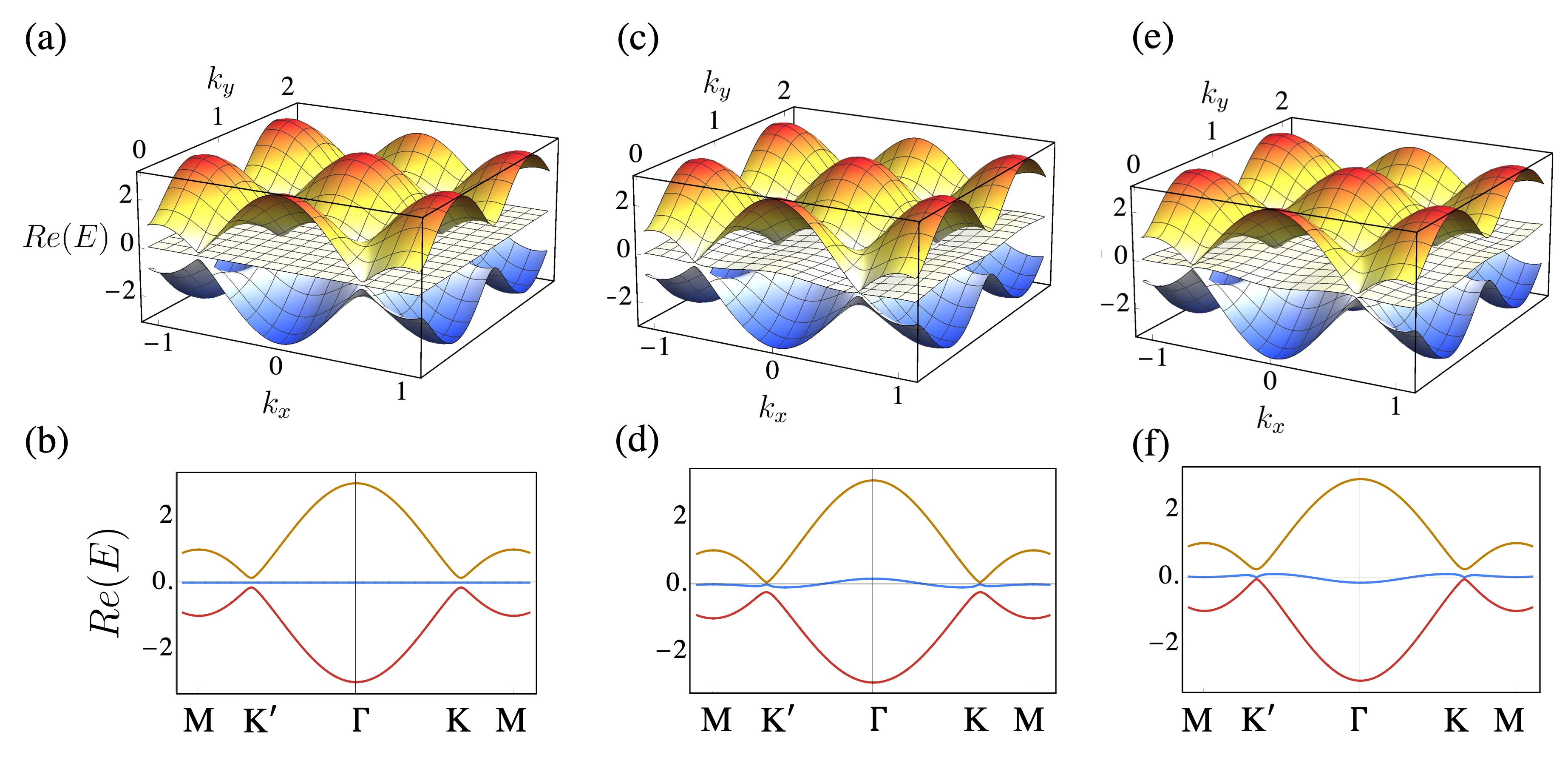}
    \caption{\label{Fig: AGVGCG} \textbf{Energy spectra of the different phases in the Hermitian model.} The band structure of the Hermitian dice-Haldane model along high symmetry points $M-K'-\Gamma-K-M$. The upper panels show the three-dimensional spectrum as a function of $k_x$ and $k_y$, while the lower panels show the corresponding two-dimensional plots at $k_y=0$. Panels (a) and (b) show the all-gapped (AG) phase with $\phi= \pi/2$ and $m=0$. Here, all three bands are gapped for all values of $k_x$ and $k_y$. Panels (c) and (d) show the valence-gapped (VG) phase where the conduction and flat bands are gapless while the valence band remains gapped. Here, $\phi=0$ and $m=0.15$. In (e) and (f) $m=0.15$ and $\phi=\pi$ which shows the conduction-gapped (CG) phase, where the conduction band is gapped while the valence and the flat bands are gapless. For all plots the values of $t=1/\sqrt{2}$ and $t_2=0.06 t$.}
\end{figure*}


We perform a Fourier transform of the Hamiltonian given in Eqn.~\ref{eq:hamadd} and Eqn.~\ref{eq:hameach} in order to obtain the energy band diagram in the two-dimensional momentum space. To encompass all the relevant physics of the model, we choose a symmetry path $ M (1,0) -K^{'} (-2/3,0)- \Gamma (0,0)-K(2/3,0)-M(1,0) $ in the BZ, which includes all the high-symmetry points. Note that the coordinates of the symmetry points are given in units of $2 \pi /a$, where $a$ is the lattice constant of the dice lattice. Later, we will systematically invoke non-Hermiticity in our model and study the physics around these high-symmetry points. 
The dice lattice with only nearest neighbour interactions ($H_{nnn} = H_{m} = 0$), exhibit two Dirac-like dispersive bands while a dispersion-less flat band lies at the Fermi level as shown in Fig. \ref{Fig: 3} (a) and (b). The Dirac points lie at the symmetry points $K$ and $K^{'}$ of the BZ. Next, the complex next nearest neighbour hopping term ($H_m = 0$) splits the Dirac cones as shown in Fig. \ref{Fig: 3} (c) and (d) resulting in non-trivial topological band structures. On the contrary, only Semenoff mass term ($H_{nnn} = 0$) induces a trivial or normal band gap in the system as given in Fig. \ref{Fig: 3} (e) and (f). It is worth noting that both $H_{nnn}$ and $H_m$ open up a gap in the energy spectrum even for arbitrarily small values of $t_2$ and $m$. The competing nature of $H_{nnn}$ and $H_m$ leads to the rich topological phase diagram of the dice-Haldane model.

In particular, Fig. \ref{Fig: 8} illustrates the effect of the Semenoff mass on the electronic band structure of the non-trivial dice-Haldane model. Even an arbitrarily small non-zero value of $m$ opens up a band gap at the $K$ and $K^{\prime}$ points of the BZ while maintaining topologically non-trivial features [Fig. \ref{Fig: 8}(a)]. With a further increase in $m$, for a critical value $m=m^*$, the $K^{\prime}$ point becomes triply degenerate while the $K$ point develops no such band touching [Fig. \ref{Fig: 8}(b)]. Beyond this point, for $m>m^*$, the system becomes topologically trivial and the energy bands are completely non-degenerate for all higher values of $m$ [Fig. \ref{Fig: 8}(c)]. It is interesting that the different phases of the Hermitian dice-Haldane model exhibit qualitatively different energy band structures. The AG phase features three non-degenerate bands as shown in Fig. \ref{Fig: AGVGCG} (a) and (b). The upper panel in Fig. \ref{Fig: AGVGCG} shows the three-dimensional band structure as a function of $k_x$ and $k_y$, while the lower panel is a two-dimensional plot of the dispersion relations along line ($M-K^{'}- \Gamma-K-M$) joining the high symmetry points of the BZ. The VG phase is shown in Fig. \ref{Fig: AGVGCG} (c) and (d) where the valence band remains gapped while the flat band and conduction band are degenerate at some points in the BZ. Fig. \ref{Fig: AGVGCG} (e) and (f) show the CG phase where the conduction band is non-degenerate while the valence and flat bands become gapless.

\section{Effect of gain and loss in the dice lattice model}\label{apd2}

\begin{figure*} [hbt!]
    \centering
    \includegraphics[width=0.9\textwidth]{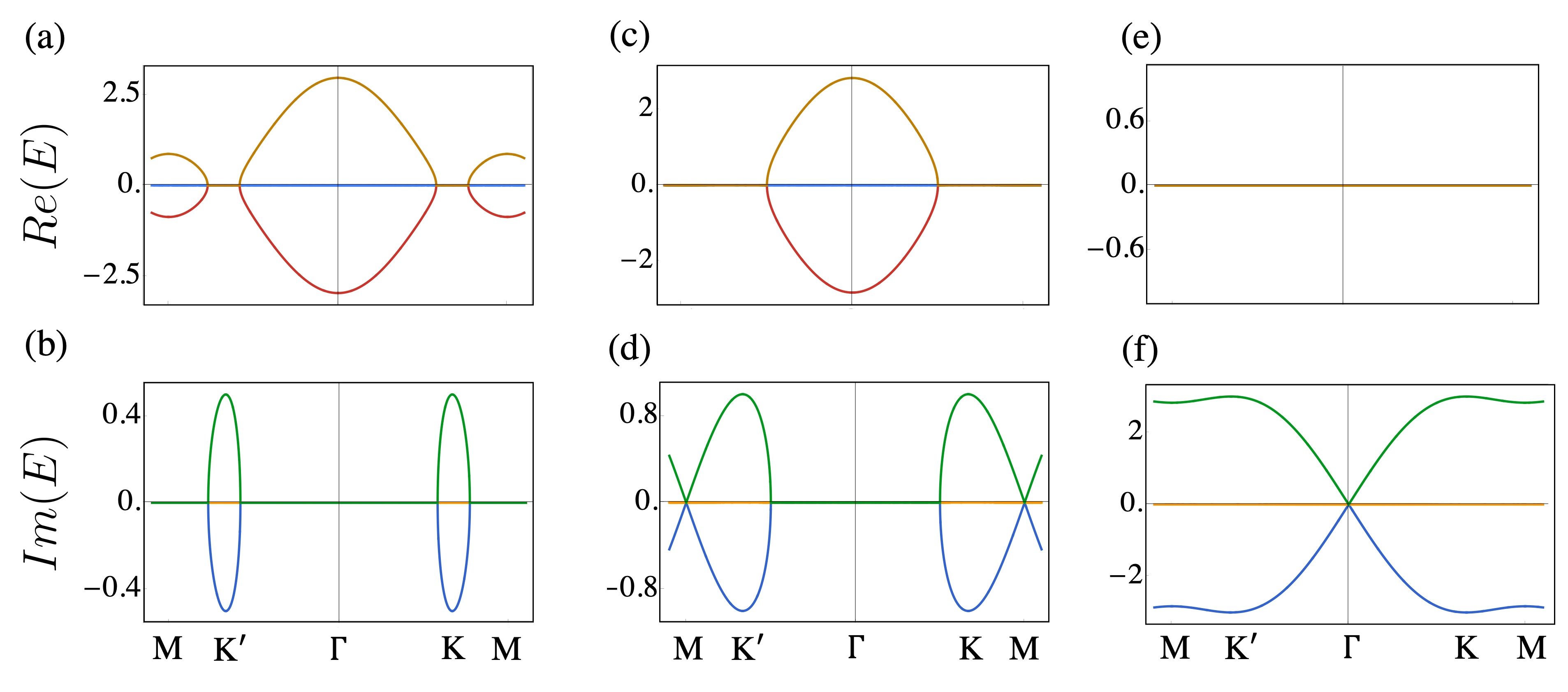}
    \caption{\label{Fig: 4} \textbf{Spectra with nearest-neighbour hopping and non-Hermiticity.} The $Re(E)$ (upper panels) and $Im(E)$ spectra (lower panels) have been shown for different values of non-Hermiticity strength $\delta$. In (a) and (b) $\delta=0.5$ showing that the non-Hermitian gain and loss instantly makes the $K$ and $K'$ points of $Re(E)$ degenerate while the degeneracy is lifted in $Im(E)$. In (c) and (d) $\delta=1.0$, this induces an EP at the $M$ point where both the $Re(E)$ and $Im(E)$ are simultaneously degenerate. Similarly in (e) and (f) we find an EP at the $\Gamma$ point for $\delta=3.0$. For all plots we have set $t=1/\sqrt{2}$.}
\end{figure*}

\begin{figure}[hbt!]
    \centering
    \includegraphics[width=0.33\textwidth]{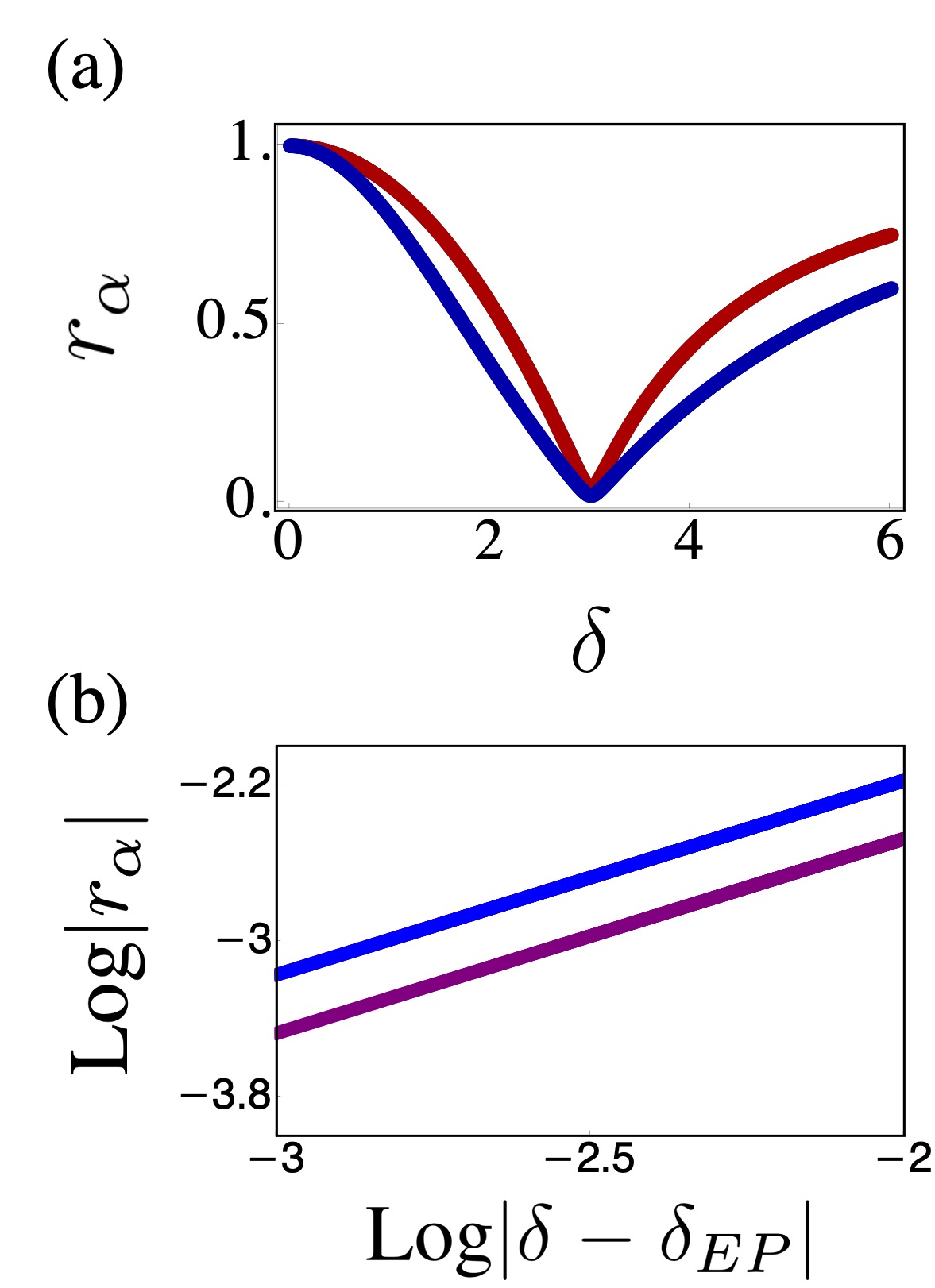}
    \caption{\label{Fig: 5} \textbf{Phase rigidity for nearest-neighbour hopping with non-Hermiticity.} Panel (a) shows the phase rigidity of the eigenfunctions $r_{\alpha}$ as a function of non-Hermiticity $\delta$, while panel (b) shows the scaling of the corresponding $r_{\alpha}$ around the concerned EP. Here, the plots have been shown for the $\Gamma$ point where an EP is induced at $\delta=3.0$. At this point, $r_{\alpha} \rightarrow 0$ shown in (a). Panel (b) shows the logarithmic scaling of $r_{\alpha}$ around $\delta_{EP}=3.0$ gives a slope of unity implying it is an EP of order three. The different colours in both the plots correspond to different eigenstates. The $r_{\alpha}$ for the two dispersive bands overlap. For both plots we have set $t=1/\sqrt{2}$.}
\end{figure} 

We consider the nearest neighbour hopping in the Hamiltonian, while keeping $t_2$ and $m$ switched off. We introduce non-Hermiticity to the system and study its effects as we vary the strength of gain and loss, $\delta$. In this condition, the Hamiltonian has the form $H = H_{nn} +H_\delta$, where $\delta$ is tuned methodically. The band diagram for the Hermitian case, i.e., at $\delta = 0$, has been previously shown in Fig. \ref{Fig: 3}(a) and (b). As we invoke gain and loss, a complex dispersion relation appears even for arbitrarily small values of $\delta$. Such a dispersion relation results in a complex eigenvalue spectrum owing to the non-Hermitian nature of the Hamiltonian. Particularly, the real part of the spectra in this condition become a single-sheeted hyperboloid around the $K$ point, near which the imaginary part of the spectra has a non-vanishing contribution as shown in Fig. \ref{Fig: 4}(a) and (b). Here, the appearance of complex energy spectra is exciting because the $PT$ operator still commutes with the Hamiltonian. Therefore, the only explanation is that the Hamiltonian and $PT$ symmetry operators do not possess the same set of eigenvectors underpinned by the anti-linear nature of the $T$ operator. Similar features have also been evinced for graphene in the presence of non-Hermitian gain and loss\cite{kremer2019demonstration}. This can be explicitly shown for the dice lattice where the $\hat{P}$ operator maps A $\leftrightarrow$ C lattice sites and is given by the following matrix:

\begin{equation}
  \hat{P}= \begin{pmatrix}
  0 & 0 & 1\\
  0 & 1 & 0\\
  1 & 0 & 0
  \end{pmatrix},
\end{equation}

and $\hat{T}$ is the antiunitary complex conjugation operator for our spinless system. It can be shown that the commutation relation $[\hat{P}\hat{T},H(k)]$ is invariably zero for arbitrary values of $\delta$. On the other hand, the eigenvectors of $\hat{P}\hat{T}$ and $H$ operators, particularly shown below for the $\Gamma$ point, are clearly distinct.

\begin{equation}
  \hat{P}\hat{T} \rightarrow \begin{bmatrix}
 -1 \\ 0 \\ 1
  \end{bmatrix},
  \begin{bmatrix}
  1 \\ 0 \\ 1
  \end{bmatrix},
  \begin{bmatrix}
 0 \\ 1 \\ 0
  \end{bmatrix}.
\end{equation}

\begin{equation}
  H \rightarrow \begin{bmatrix}
 -1 \\ a \\ 1
  \end{bmatrix},
  \begin{bmatrix}
  a\eta_- \\ \eta_-\\ 1
  \end{bmatrix},
  \begin{bmatrix}
 a\eta_+\\ \eta_+ \\ 1
  \end{bmatrix},
\end{equation}

where $\eta_{\pm}= a \pm \sqrt{2+a^2}$ and $a=\frac{\sqrt{2}}{3} i \delta$.
Hence, despite $PT$ symmetry of the system, the eigenstates correspond to the broken-$PT$ phase.\\

A further increase in $\delta$ extends the degeneracy of $Re(E)$ to $M$, while the degeneracy at this point in $Im(E)$ is lifted as presented in Fig. \ref{Fig: 4}(c) and (d). Most interestingly, this critical transition at the $M$ point occurs exactly at $\delta=1.0$. At this value of $\delta$, we find that the three eigenvalues and eigenfunctions coalesce at the $M$ point, giving rise to a third order EP. As we increase $\delta$ further, the branches of $Re(E)$ becomes degenerate while the degeneracy is lifted in $Im(E)$ at the corresponding $k_x$ values. At $\delta=3.0$, we find the coalescence of the three eigenvalues and eigenfunctions at the $\Gamma$ point [Fig. \ref{Fig: 4}(e) and (f)]. This is again a third order EP, however appearing at a different point in the BZ. The phase rigidity and its scaling for this EP are shown in Fig. \ref{Fig: 5}. Fig. \ref{Fig: 5}(a) shows the variation of phase rigidity $r_\alpha$ of the eigenvectors as a function of $\delta$. $r_\alpha \rightarrow 0$ as $\delta \rightarrow \delta_{EP}=3.0$. The scaling of the phase rigidity has been plotted on a logarithmic scale in Fig. \ref{Fig: 5}(b) whose slope is one, denoting that the higher order EP is indeed of order three.

\section{Non-Hermiticity and Edge States in the dice-Haldane nanoribbon}\label{apd3}

\begin{figure}[hbt!]
    \centering
    \includegraphics[width=0.45\textwidth]{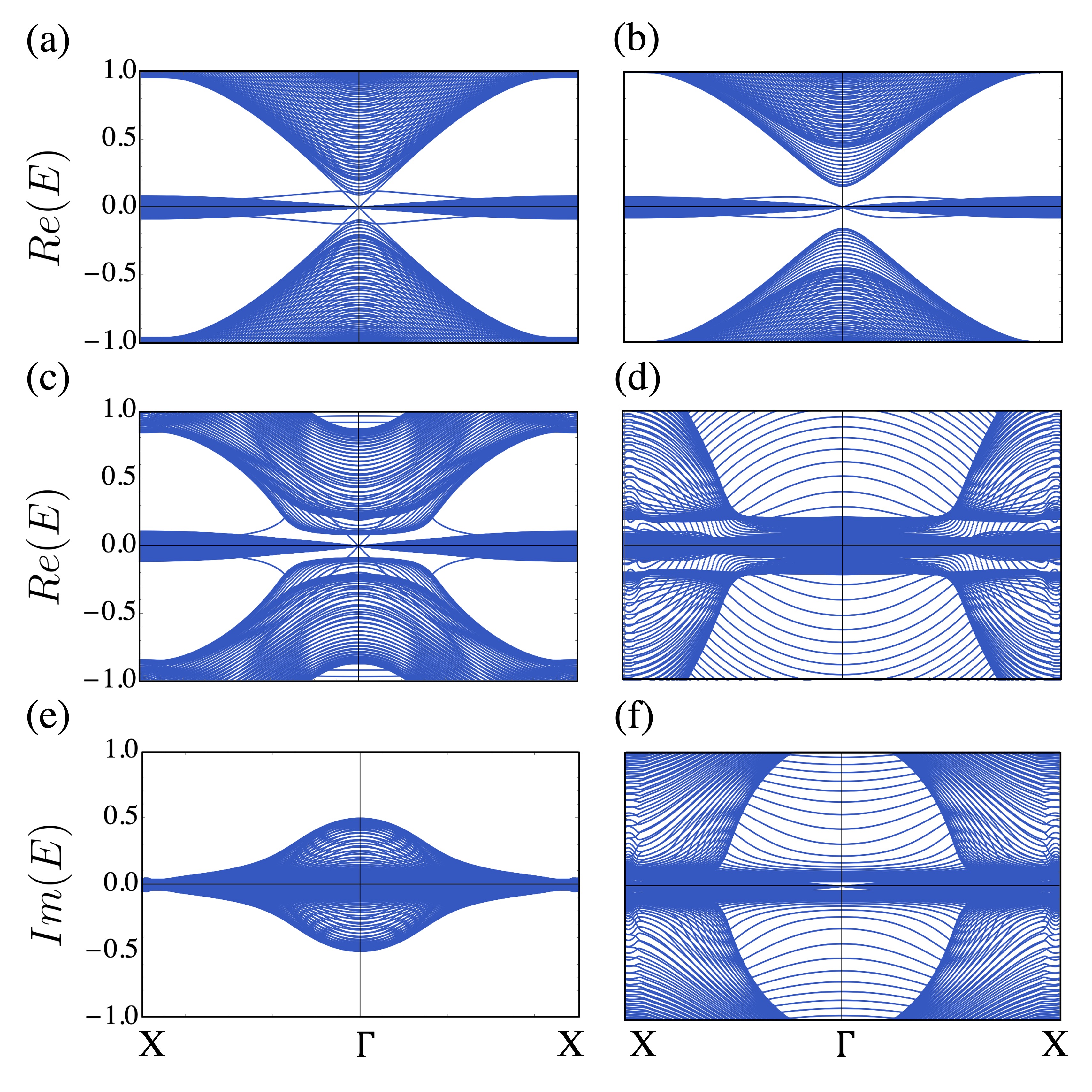}
    \caption{\label{Fig: Ribbon_gainloss} \textbf{Effect of gain and loss on the edge states of the nanoribbon.} Here, the nanoribbon has been considered to be finite in the $y$ direction while we have imposed PBC along $x$. Panels (a) and (b) correspond to the Hermitian nanoribbon. Panel (a) shows the real spectrum for $m=0.06$ and panel (b) shows the spectrum for $m=0.3$. The spectra have been plotted along high symmetry lines $X-\Gamma-X$. In the topologically non-trivial region we find conducting edge states in the spectrum [shown in panel (a)], while in the topologically trivial region [shown in panel (b)] no non-trivial edge states are present. Panels (c) and (d) show the persistence of edge states in the nanoribbon with non-Hermitian gain and loss while in the topologically non-trivial region ($m=0.06$). The edge states are robust to $\delta$ up to a critical value of $\delta_c \approx 1$, after which they disappear. (c) The presence of edge states in the real energy spectra can be seen for $\delta=0.5 <\delta_c$. (d) The edge state is absent for $\delta=1.5> \delta_c$, as expected. The corresponding imaginary spectra for (c) and (d) have been shown in (e) and (f) respectively. The system with non-Hermiticity has a finite non-zero $Im(E)$ yet the real part of the spectra can still accommodate edge states. Here, the number of hexagonal layers in the $y$ direction has been taken to be 70. We have chosen $t=1/\sqrt{2}$, $t_2=0.06 t$, and $\phi=\pi/2$.}
\end{figure}

Topological zero energy modes which are found in the non-trivial region of the Hermitian dice-Haldane model have been shown in Fig. \ref{Fig: Ribbon_gainloss} (a), characterised by the linear zero-energy crossings from the conduction to the valence band. The zero modes dissapear as one moves into the topologically trivial region of the phase diagram, as shown in Fig. \ref{Fig: Ribbon_gainloss}(b). Fig. \ref{Fig: Ribbon_gainloss}(a) and (b) show the spectrum close to the Fermi level ($E=0$) along the line joining high symmetry points: $X-\Gamma-X$.

Sitting in the topologically non-trivial region, we have studied the energy spectrum for a varying non-Hermiticity strength, $\delta$. We found that the existing edge states are robust to values of $\delta$ up to $\delta_c=0.8$ for the energy scale of our system, after which the edge states cannot be discerned from the bulk states. This can be seen in Fig.~\ref{Fig: Ribbon_gainloss}(c), where we still decipher clear edge states for $\delta<\delta_c$ and subsequently the edge states disappear for $\delta>\delta_c$ [Fig. \ref{Fig: Ribbon_gainloss}(d)]. Fig.~\ref{Fig: Ribbon_gainloss}(e) and (f) show the imaginary spectra corresponding to Fig.~\ref{Fig: Ribbon_gainloss}(c) and (d) respectively. We can see that the spectra develops finite non-zero $Im(E)$ as a result of introducing non-Hermiticity into the nanoribbon, yet the real spectra can accommodate topological edge states upto $\delta_c$. 

\begin{figure}[hbt!]
    \centering
    \includegraphics[width=0.45\textwidth]{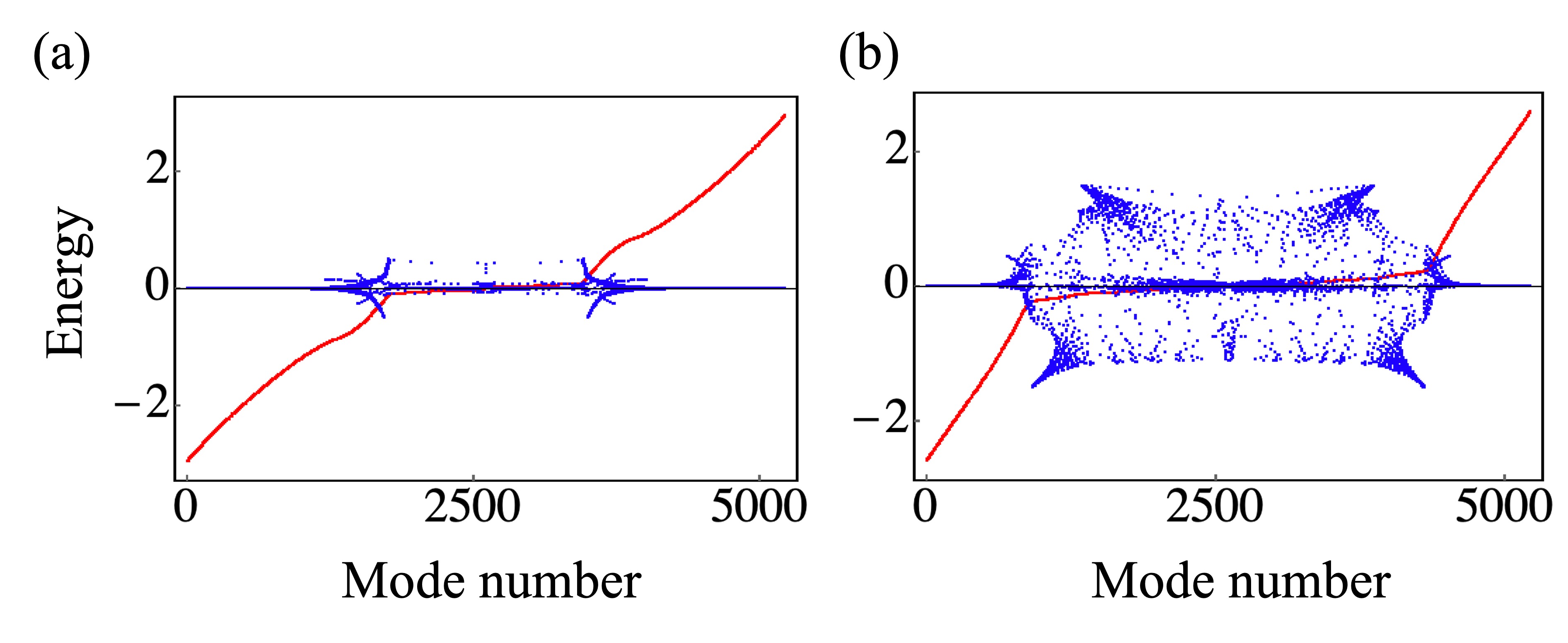}
    \caption{\label{Fig: Energies} \textbf{Real and imaginary parts of energy for the dice-Haldane nanoribbon with gain and loss.} $Re(E)$ is shown in red and $Im(E)$ is shown in blue. Panel (a) shows $Re(E)$ and $Im(E)$ for gain and loss strength $\delta = 0.5$. Panel (b) corresponds to $\delta=1.5$. The other parameters for the plots are $t=1/\sqrt{2}$, $t_2=0.06 t$, $m=0.0$, and $\phi=\pi/2$. The total number of lattice sites $n=5220$. The number of edge states that remain in the dissipation- or amplification-free region is reduced with increasing gain and loss strength.}
\end{figure}

It is well established that the bulk boundary correspondence in a system generally breaks down due to the presence of non-Hermiticity~ \cite{helbig2019observation,xiong2018does,lee2016anomalous,xiao2020non,bergholtz2021exceptional}. As a consequence, the Chern number of the momentum space bulk Hamiltonian fails to predict the existence of topological edge states correctly. In principle, a topological invariant may be calculated by taking into account the GBZ formalism~\cite{yang2020non}. However, for a two-dimensional, non-trivial model, it is a challenging task to deduce the GBZ. An alternative approach has been explored in Ref.~\cite{resendiz2020topological}, where a finite two-dimensional topological Haldane lattice has been constructed with arbitrary edge types. In these finite two-dimensional systems, edge states appear even if the eigenvalues are not entirely real due to the presence of gain and loss. Moreover, in this $\mathcal{PT}$ symmetry broken phase, the topological protection has been determined by the number of edge states that remain within the dissipation- or amplification-free region. The absence of back-scattering, hence the topological phase, has been confirmed by calculating the time evolution of an edge state. As expected, the number of real edge states essentially depends on the strength of gain and loss ($\delta$), which drives the topological transition. We have used a similar formalism for our dice-Haldane model to calculate the critical values of $\delta$ ($\delta_c$) corresponding to the topological protection. The numerical value of $\delta_c$ has been obtained from the maximum value of gain or loss for which at least 5\% of the edge states remain dissipation- or amplification-free. We note that the value of $\delta_c$ for the dice-Haldane system varies with system size similar to the conventional Haldane model. For example, $\delta_c = 0.8$ for a smaller lattice size, say $n= 288$. However, for the system with sufficiently large number of lattice sites (checked for $n = 1152$ and $n = 5220$), the critical value of $\delta_c$ saturates to $\approx 1$ as depicted in the Fig.~\ref{Fig: Energies}. Fig.\ref{Fig: Ribbon_gainloss}(c) and (d) illustrate the effect of gain and loss strength on the real part of the dice-Haldane nanoribbon band structure in momentum space. Here, under the time-reversal symmetry broken condition, the Semenoff mass term is chosen as $m= 0.06$, ensuring the non-trivial topological phase of the Hermitian system to start with. We observed that the non-Hermiticity in terms of gain and loss above the critical value essentially destroys the topologically protected edge states in the dice-Haldane lattice.

\section{Geometry dependent skin effect under non-Hermitian gain and loss}\label{apd4}

The finite spectral area of the dice-Haldane torus (PBC in both directions) in the complex plane cannot solely explain the appearance of skin effect only on the top and bottom edges. The nature of skin effect shown in Fig. \ref{Fig: GL}(b) and (d) under non-Hermitian gain and loss resembles the geometry-dependent skin effect mentioned in Ref. \cite{Zhang_2022}. To illustrate, we calculate the Zak phase~\cite{zhang2019partial} and the corresponding winding number ($W$) of the filled bands~\cite{banerjee2022chiral} along different directions, supporting our findings of the skin effect. Additionally, we elucidate how the complex energy spectrum can also be used to predict NHSE in the top and bottom edges ($y$-direction) in Fig. \ref{Fig: GL}(b) and (d) rather than along the length ($x$-direction), which can occur on imposing non-reciprocal hopping [Fig. \ref{Fig: NRH}]. For that purpose, we consider the following two cases --  

\begin{figure}[hbt!]
    \centering
    \includegraphics[width=0.45\textwidth]{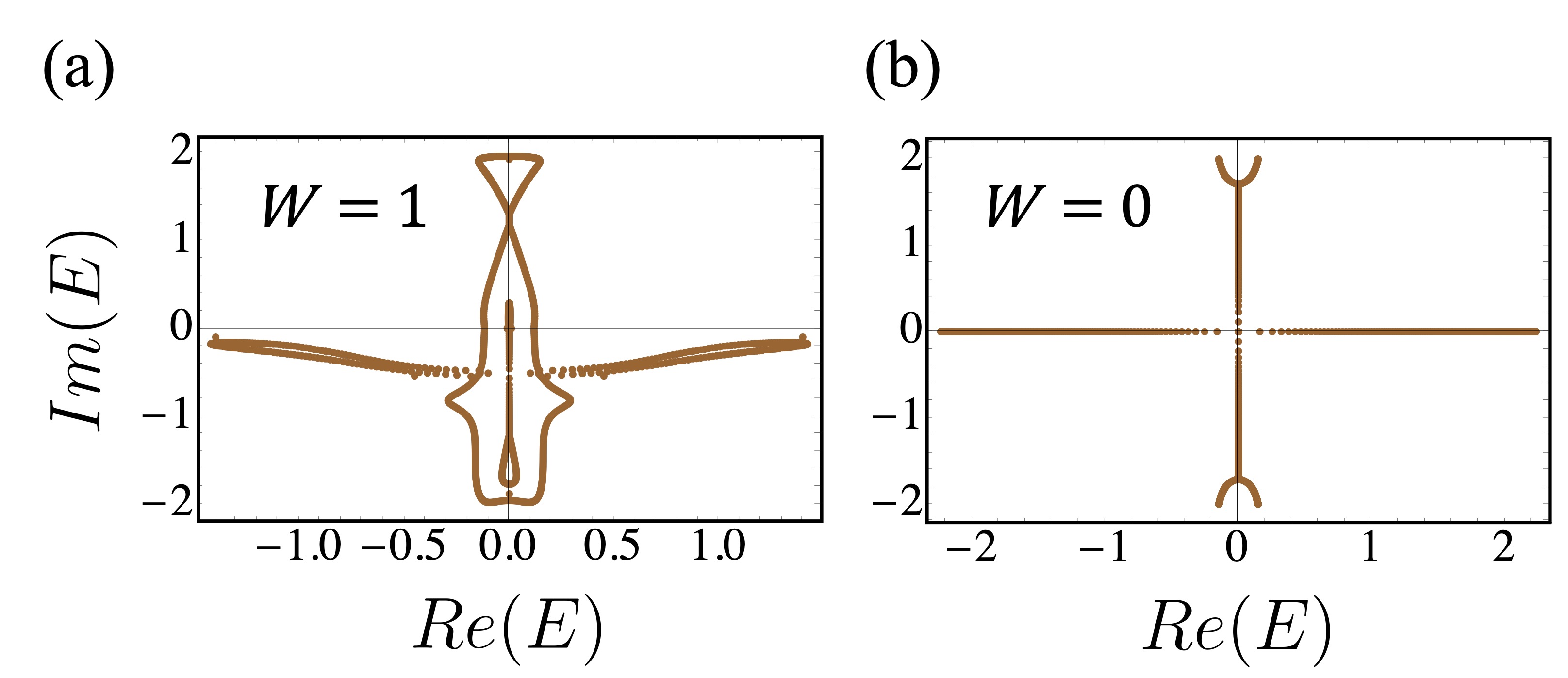}
    \caption{\label{Fig: spectrum} \textbf{Complex energy spectra and winding number with gain and loss in the dice-Haldane ribbon under two different orientations.} Panel (a) shows the complex energy spectrum of the ribbon with PBC along $y$ and OBC along $x$. The spectrum encloses a finite spectral area which implies the existence of a skin effect in the $y$ direction when one imposes OBC. The skin effect has been confirmed by the winding number, $W=1$. Panel (b) shows the complex spectrum of the ribbon with PBC along $x$ and OBC along $y$. The spectrum shows an arc-like structure with no enclosed area. This signifies that no skin effect will occur along the $x$ direction when OBC is imposed. The absence of skin effect in this case has been confirmed by the winding number, $W=0$. For the above plots the values of $t=1/\sqrt{2}$, $t_2=0.06 t$, $m=0$, $\delta=2.0$, and $\phi=\pi/2$.}
\end{figure}

Case1: We impose PBC along the $y$ direction while we have OBC along $x$. The energy spectrum in the complex plane of this nanoribbon under gain and loss shows a loop-like structure enclosing a finite spectral area. This implies a skin effect when OBC is imposed along the $y$ direction, i.e., along the top and bottom edges of the sheet. To confirm this observation, we have calculated the Zak phase and the corresponding winding number. We find that the filled bands contribute to $W=1$ [See: Fig.~\ref{Fig: spectrum}(a)] as per expectation.   

Case 2: We consider the ribbon to be periodic in the $x$ direction and open along $y$. The spectrum in this case has an arc-like structure in the complex plane, which does not enclose a finite spectral area. This implies that on imposing OBC in the $x$ direction, skin effect will not occur at the right or left edges. In this geometry we find $W=0$ for the filled bands. [See: Fig.~\ref{Fig: spectrum}(b)].

\def\urlprefix{}
\def\url#1{}
\bibliographystyle{apsrev}
\bibliography{Ref}


\end{document}